\documentclass[aps,final,notitlepage,oneside,twocolumn,superscriptaddress,showpacs,centertags]{revtex4-1}
\usepackage{graphicx}
\usepackage{latexsym}
\usepackage{amssymb}
\usepackage{amsmath}
\usepackage{psfrag}
\usepackage{float}

\def\be{\begin{equation}}
\def\ee{\end{equation}}
\def\bea{\begin{eqnarray}}
\def\eea{\end{eqnarray}}
\def\ba{\begin{array}}
\def\ea{\end{array}}
\def\bse{\begin{subequations}}

\def\m{{\mu}}
\def\n{{\nu}}
\def\d{{\delta}}
\def\e{{\epsilon}}
\def\L{{\mathcal{L}}}

\def\cA{{\cal A}}

\begin{document}

\title{Black holes in the presence of dark energy}
\author{E. O. Babichev}\thanks{e-mail: eugeny.babichev@th.u-psud.fr}
\affiliation{Laboratoire de Physique Theorique d'Orsay CNRS UMR 8627, \\
Universit\'{e} Paris-Sud 11 91405 Orsay Cedex, France}
\author{V. I. Dokuchaev}\thanks{e-mail: dokuchaev@inr.ac.ru}
\author{Yu. N. Eroshenko}\thanks{e-mail: eroshenko@inr.ac.ru}
\affiliation{Institute for Nuclear Research, Russian Academy of Sciences \\
prosp. 60-letiya Oktyabrya 7a, 117312 Moscow, Russian Federation}


\begin{abstract}
The new, rapidly developing field of theoretical research --- studies of dark energy interacting with black holes (and, in particular, accreting onto black holes) --- is reviewed. The term `dark energy' is meant to cover a wide range of field theory models, as well as perfect fluids with various equations of state, including cosmological dark energy. Various accretion models are analyzed in terms of the simplest test field approximation or by allowing back reaction on the black-hole metric. The behavior of various types of dark energy in the vicinity of Schwarzschild and electrically charged black holes is examined. Nontrivial effects due to the presence of dark energy in the black hole vicinity are discussed. In particular, a physical explanation is given of why the black hole mass decreases when phantom energy is being accreted, a process in which the basic energy conditions of the famous theorem of nondecreasing horizon area in classical black holes are violated. The theoretical possibility of a signal escaping from beneath the black hole event horizon is discussed for a number of dark energy models. Finally, the violation of the laws of thermodynamics by black holes in the presence of noncanonical fields is considered.
\end{abstract}


\maketitle

\tableofcontents


\section{Introduction}

The discovery of accelerating expansion of the universe is one of the most important cosmological discoveries at the turn of the 20th and 21st centuries \cite{nob1,nob2,nob3}. Independent evidence of the accelerating expansion has been obtained from type-Ia supernova observations, from measurements of cosmic microwave background fluctuations (integrated Sachs--Wolfe effect), from studies of the large-scale distribution of galaxies, and from gravitational lensing. According to the interpretation of the accelerating expansion using Einstein's General Relativity (GR) and the Friedmann cosmology, some form of matter exists with a negative pressure whose absolute value is approximately equal to the energy density (in units where the speed of light is $c=1$) \cite{acceler1,acceler2,acceler3,acceler4}. This matter, called dark energy, started dominating in the universe at redshifts $z\sim0,5-1$, and presently its contribution to the total energy density in the universe amounts to $\approx70$\%. The modern state of the dark energy problem is reviewed, for example, in \cite{Che08,Che13,LukRub08,BolEroLem12,Copeland:2006wr}.

It is important to note that the physical origin of dark energy remains unknown. The term `dark energy' simply reflects the observed properties of this matter: the word `dark' means that it is not directly seen in any observations except gravitational measurements, and `energy' reflects the
fact that this matter has an energy-momentum tensor that can be found from Friedmann's equations. An important feature not encrypted in the term `dark energy', is its negative pressure, which absolute value is comparable to the energy density.

Whatever the nature of dark energy, it can be effectively characterized by the pressure and density, and it is possible to introduce their ratio $w\equiv p/\rho$, also known as the equation-of-state parameter. The notion of the cosmological constant ($\Lambda$-term), which was introduced by Einstein \cite{ein13}) and subsequently rejected by him (see \cite{ein14,ein15}) after reading Fiedmann's paper \cite{fri16} (also see \cite{fri16b}) and recognizing the observational evidence of the expansion of the universe \cite{Cherep18}, has seen a rebirth due to astronomical observations since the discovery of dark energy.

Presently, the ever-growing list of proposed models for dark energy is quite long and includes very different ideas (see, e.\,g., a review of theoretical models in \cite{YooWat12}. The introduction of the cosmological $\Lambda$ term requires an extremely small value of the energy density of the vacuum (as presently observed), which demands a huge fine-tuning of field theory parameters \cite{Bur10}. For this reason, instead of introducing the $\Lambda$ term, models of dynamical dark energy with  $w\neq-1$ have been proposed. Here, for example, we can mention the popular models of a scalar field with a flat potential (the `quintessence') \cite{CaDaSt1,CaDaSt2,CaDaSt3,CaDaSt4,CaDaSt5,CaDaSt6,CaDaSt7} and models with a nontrivial kinetic term (the `$k$-essence') \cite{ArMuSt1,ArMuSt2,ArMuSt3}, the ghost condensate \cite{ghost} and Galileons \cite{Nicolis:2008in,Deffayet:2009wt,Deffayet:2009mn,Chow:2009fm}. Different generalizations and modifications of GR have been discussed in which dark energy effectively emerges (so-called geometrical dark energy) \cite{phcosm1,Maia04}. In this connection, we can mention scalar-tensor theories \cite{Jordan-1,Jordan-2,Brans:1961sx,Damour:1992we}, including $f(R)$-gravity (see, e.g., review \cite{Sotiriou:2008rp}), multidimensional models (see review \cite{Rubakov:2001kp}), and massive gravity \cite{DeFelice:2013bxa,RubTin08}. In addition, models of matter that would have the properties of dark energy at large scales and imitate dark matter (hidden mass) at small scales \cite{PadCho02} have been proposed. Models have been considered in which the apparent accelerating expansion of the universe results from averaging of small-scale density inhomogeneities with back reaction in the background metrics \cite{Claetal11}. The explanation of dark energy in terms of a gravitational wave background has also been proposed \cite{BieHar13}.

Interestingly, the modern observational data do not exclude the possibility that dark energy represents the so-called phantom energy \cite{Caldw1,Caldw2}: a matter with the effective equation-of-state parameter $w<-1$. Indeed, results from the Planck space mission in combination with polarization measurements by the WMAP (Wilkinson Microwave Anisotropy Probe) satellite and acoustic oscillations yield the parameter $w=-1.13^{+0.23}_{-0.25}$ \cite{Ade:2013zuv}, with the mean value slightly smaller than $-1$.

Different aspects of phantom cosmology have been considered in many papers (see, e.\,g., \cite{Copeland:2006wr,phcosm1}). The possibility of a Big Rip is one extravagant scenario of phantom
cosmology \cite{Caldw1,Caldw2}. In this scenario, the cosmological density of phantom energy and the scale factor of the universe diverge to infinity in a finite time interval, such that all bound objects at the scales of the effective description of phantom energy are disrupted.

We emphasize, however, that the simplest models of phantom energy turn out to be unstable, notably, due to the presence of ghosts, which lead to vacuum instability. An infinitely rapid instability can be rendered finite in time by introducing an ultraviolet (UV) cut-off \cite{CarHofTro03,CliJeoMoo04}, which, however, requires violating the Lorentz invariance of the
theory. In more complicated field models, it is possible to reach stability at least during some period of the cosmological evolution \cite{Rub06,Libetal07}. In the Galileon model, the phantom
equation of state is not something special: due to the kinetic coupling of the graviton to a scalar field inherent in this model (the mixing of kinetic terms), it is quite straightforward to obtain a stable phantom regime \cite{Deffayet:2010qz}.

The evolution of dark energy is usually considered in the context of cosmological problems, in which a homogeneous dark energy determines the cosmological expansion dynamics. However, it is also interesting to study the behavior of dark energy and different kinds of matter in general in the vicinity of black holes (BHs). Modern astronomical observations by large ground-based and space telescopes provide compelling evidence that supermassive black holes exist almost in each massive (structured) galaxy \cite{Vol10}. Possibly, primordial BHs formed at pre-galactic stages served as seeds for BHs in galactic nuclei. A model was also proposed in which the Hubble flow itself can be treated as a gravitational collapse into a black hole inverted in time, i.\,e., the universe in this model is considered as the internal part of a white hole \cite{LukMihStro12}. There is almost no doubt that supermassive BHs in galactic nuclei and stellar-mass BHs, which are remnants of stellar evolution, do exist, and it is therefore the right time to investigate the properties of these BHs and to study their interaction with different kinds of matter and fields, with dark energy in particular.

In this review, we focus on the problem of accretion of dark energy (and, in general, different kinds of matter) onto BHs. We note that the accretion rate of dark energy onto astrophysical BHs in many cases is much smaller than that of ordinary baryonic matter. Therefore, from the astrophysical
point of view, the problem of accretion of dark energy seems to be purely academic. On the other hand, BHs are extremely valuable objects from the fundamental perspective because, in some sense, they provide a test area to study different kinds of matter. In this connection, the problem of dark energy accretion becomes relevant, because in some cases it allows obtaining exact solutions and, more importantly, studying various physical effects generated by different kinds of matter
in the gravitational field of a BH.

The history of the study of accretion of a perfect fluid onto a compact star began with the pioneering work by Hoyle and Littelton \cite{HoyLyt39} and by Bondi and Hoyle \cite{BonHoy44}. In these papers, regions located far away from the BH horizon were considered, and therefore the problem could be treated nonrelativistically. Later, the problem of accretion in the Newtonian limit was solved in the classical paper by Bondi \cite{Bon52}. Stability of the Bondi accretion with respect to small perturbations was studied in \cite{Gai06}. The generalization to the case of accretion of a relativistic gas was done by Michel \cite{Mic72} ((see also additions to the Michel solution in \cite{Beg2,Beg3,Beg6,Beg7,PetShaTeu} and details of the history of the accretion theory in \cite{Cha04}).

We emphasize that in the accretion papers cited above, the infalling matter is treated as a test
fluid, i.\,e., the back reaction of the fluid on the metric is usually neglected. However,
when, for example, considering the formation of primordial BHs in the universe, such an approximation is insufficient, and the full system of equations must be solved. The problem of primordial BH formation was first formulated by Zeldovich and Novikov \cite{ZN}. Later on, Carr and Hawking \cite{Beg1} considered the problem of accretion of dust and radiation onto a primordial BH immediately after its birth and at later stages. Carr and Hawking solved the full system of Einstein's equations, taking the back reaction of the accreting fluid into account. This idea has been further elaborated in many papers (see, e.\,g., \cite{accretion1,accretion2,accretion3,accretion4,accretion5,accretion6,Carr:2010wk,Beg5,Mal99}. In particular, the dynamics of the horizon during collapse and accretion was studied in \cite{ShaAnd99}.

Models of accretion onto astrophysical BHs (from accretion disks in particular), taking magnetic fields and the realistic thermodynamics of matter into account, have been considered in many papers (see, e.\,g., reviews \cite{BesPar93,Bes97,Bes03} and monograph \cite{Bes05}).

In this review, we do not discuss problems of magnetic hydrodynamics. We consider classes of problems in which the matter equation of state can be somewhat simplified and idealized. Moreover, we assume spherically symmetric accretion in almost all of our calculations. These assumptions
allow us to obtain exact solutions and to address fundamental questions on the internal structure and fate of black holes.

The generalization of the Bondi--Michel accretion to dark energy was proposed in \cite{BabDokEro04,BabDokEro05-2}. In these papers, dark energy was modeled by a perfect fluid with the equation of state $p=p(\rho)$, and the problem of quasispherical accretion onto BHs was considered. In particular, in \cite{BabDokEro04}, accretion onto a BH in a universe filled with evolving phantom energy, when dark energy determines both the dynamics of the expanding universe and the evolution of the accreting BH, was studied.

In \cite{accretion5,accretion6,Carr:2010wk}, self-similar solutions for a BH on a cosmological background are discussed, and the question is addressed of whether the BH growth rate can be equal to that of the cosmological horizon.

Accretion of dark energy onto realistic astrophysical BHs (intermediate-mass BHs in globular clusters) was discussed in \cite{PepPelRom11}, and the conclusion was made that accretion of dark
energy has no observational consequences in this case.

We note that the problem of accretion of a perfect fluid can be reformulated (under some assumptions) in terms of a scalar field with shift symmetry. However, a scalar field with a nontrivial potential cannot be described by a perfect fluid. Therefore, it makes sense to investigate accretion of a scalar field, because different scalar fields have been proposed as
dark energy candidates.

In \cite{Jac,FroKof,Unruh1,Unruh2,Unruh3,CruGuzLor11,GuzLor12}, the behavior of a scalar field with the canonical kinetic term near a BH was studied for different potentials $V(\phi)$. and some analytic solutions for the BH mass evolution were obtained. It was shown in \cite{Fro04} that accretion of ghost condensates (fields of a special kind) onto a BH can be very effective (however, see the criticism of this approach in \cite{Mukohu}).

It is the study of different kinds of matter near BHs (which in many cases first appears as dark energy) that often yields interesting and unexpected results, which we discuss in this review. A decrease in the BH mass due to the phantom energy accretion is one such result \cite{BabDokEro04,BabDokEro05-2,ForRom01} (also see \cite{Sha07}). The BH mass decreases because the perfect fluid energy flux is proportional to $\rho+p$, which is negative by definition for phantom energy (see, e.g., \cite{LL8}). In a universe filled with phantom energy, the masses of all BHs gradually vanish as the evolution approaches the Big Rip \cite{BabDokEro04}. This decrease in BH masses is due to violation of the weak energy condition $\rho+p\geq0$, which underlies theorems on the nondecreasing surface of classical BHs (ignoring quantum effects) \cite{hawkell}. The conclusion that accretion of a scalar field with nonminimal coupling, which violates the energy conditions, leads to the BH mass decreasing was previously formulated in \cite{ForRom01} (also see \cite{RodSaa09}). The decrease in the BH horizon during accretion of a phantom scalar field is confirmed by numerical calculations in \cite{GonGuz09,LorGonGuz12}, which also showed that this decrease is not an artefact of the reference frame choice.

Recently, the possibility of accretion of an exotic fluid with negative density, whose existence is not fully excluded by GR, has also been discussed \cite{Sch10,Bon89,ShaNovKar11,Iva12}. In the real world, such fluids may correspond to some quantum systems, for example, to the Casimir energy.

Hypothetical microscopic BHs, which can arise due to quantum gravity effects, represent another limit case, which is opposite to the supermassive BHs in galactic nuclei. Microscopic BHs have also been discussed from the practical point of view in relation to their hypothetical creation in the
largest accelerator experiments in models of gravity with extra dimensions.

The electric charge can be essential for microscopic BHs. Studies of charged BHs are important to clarify the key points of the theory of gravitation. In particular, it is interesting to study the features of accretion onto charged BHs and the character of the space-time changes during accretion onto such BHs. Studies of charged BHs are also of interest from the point of view of the existence of extreme BHs, which in some sense can be considered an intermediate case between black holes and `naked' singularities. We note that the extreme state of a charged BH can also be attained in a finite time interval during accretion of phantom energy if the fluid is treated as a test liquid \cite{MadGon08,JamRasQad08,BabDokEro11}. However, the back reaction of the gravity of dark energy on the metric can prevent the BH from turning into a naked singularity, in accordance with the
third law of the BH thermodynamics \cite{bch73}.

Accretion of a phantom field onto charged BHs in the theory with a $5$-dimensional (5D) space-time was studied in \cite{ShaAbb11}. The conclusion was made that in the $5D$ case, the accreting BH cannot pass through the extreme state, and the naked singularity does not emerge. In the $4D$ case, however, it is impossible to make such an unambiguous conclusion.

If a naked singularity does appear in some physical process, it is interesting to investigate the behavior of dark energy in its vicinity. Under certain assumptions, accretion of some kinds of matter is also possible onto a naked singularity. But under conservative physical assumptions, perfect fluids cannot accrete onto a naked Reissner--Nordstr\"om singularity \cite{Babetal08,BabDokEro11}, instead, a static atmosphere emerges around the singularity. A similar result was obtained numerically in \cite{Bambi09} for a Kerr naked singularity (with angular momentum).

Accretion of some noncanonical fields provides an intriguing possibility to look inside the `usual' BH horizon \cite{BabMukVik08,Bab11}; even Lorentz-invariant scalar field theories, generally speaking, allow superluminal propagation of perturbations for nontrivial configurations when the light cone lies inside the `sound' cone. \cite{MukhGar,MukhVik}. We note that the causality property becomes very nontrivial in such theories and requires a thorough investigation. For example,
the Cauchy problem cannot be solved for arbitrary initial conditions \cite{ArmenLim,Rendall,ArkHamDubov,HashItz}. The presence of such `superluminal' fields in the gravitational field of a BH opens up the possibility of `looking inside' the BH.

As mentioned above, the usual analytic treatment of the accretion problem assumes the test character of the accreting fluid. However, it is of great interest to study the back reaction of the fluid on the metric. This is a very complicated problem, however: only a few analytic solutions are known
that take the back reaction into account. The famous Tolman solution for dust accretion onto a BH \cite{Lemaitre1,Lemaitre2}, as well as the Vaidya solution \cite{Vaidya1,Vaidya2,Vaidya3,Vaidya4}, which describes a BH emerging in radially moving radiation, provide examples. There is a
generalization of the Vaidya solution that involves a more general energy-momentum tensor (see, e.\,g., \cite{Wang:1998qx}).

Another approach to the problem is also possible: instead of solving the exact back-reaction problem, one can use perturbation theory methods. In this way, for example, in  \cite{Yor85}, corrections to the metric of a Hawking-evaporating BH due to the gravitational field of the outgoing radiation were calculated. In \cite{BabDokEro12,DokEro11}, for the accretion of matter with an arbitrary equation of state, small back-reaction corrections to the metric were found. Although this method does not allow calculating large corrections, it gives fairly general results and allows finding conditions when the back reaction prevents the use of the formalism of successive approximations. The method of thin self-gravitating shells provides another useful approach. This method was formulated by Israel \cite{Isr66,Isr66b}, and has been elaborated in many papers (see review \cite{BerKuzTka} and the references therein). It turns out that his method can also
be used to model the accretion of phantom energy onto BHs.


\section{Stationary accretion}
\label{idealsec}

In this section, we consider the simplest case of spherically symmetric stationary accretion of dark energy modeled as a perfect fluid. The fluid is treated as a test flow, i.\,e., it moves in
the given external gravitational field and its own gravitational field can be neglected. This condition holds for sufficiently light fluids. The stationarity assumes that the BH mass
increases slowly, such that the distribution of the fluid on the relevant space-time scales has time to adjust itself to the changing BH metric.

\subsection{Accretion in the Newtonian approximation}
\label{BondiAccretion}

Early calculations of accretion onto the central mass were carried out in the Newtonian approximation. The test particles move in the Newtonian gravitational potential $V=GM/r$. If particles of the fluid interact weakly with each other, i.\,e., their free-path length is much larger than the characteristic scales, dust-like accretion occurs. The accretion rate is determined by the geometrical size of the central body, taking the gravitational focusing into account. If particles belong to some steady system, for example, they are stars in a globular cluster, then in the dynamical time (the time of one flight across the system) particles with small angular momentum (or, as it is said, from the `loss cone') fall onto the BH, and then the accretion rate decreases.

For a fluid, the accretion rate can be much higher than in the case of noninteracting particles, because the interaction of the fluid particles changes the directions of their momenta, and the loss cone permanently replenishes. The bulk motion velocity of the medium at infinity, v, plays an important role in accretion calculations. For noninteracting particles, this velocity and the impact parameter determine the possibility of the fall of a particle onto the BH \cite{HoyLyt39}, and the accretion rate is $\dot M\propto v^{-3}$. For a fluid, the sound speed in the medium is also important, as was shown in the classical paper by Bondi \cite{Bon52} (that is why this type of accretion is called the `Bondi accretion').

The rate of spherically symmetric stationary accretion of a  fluid with a polytropic equation of state is calculated from the solution of the Bernoulli equation
\begin{equation}
-\frac{GM}{r}+\frac{v^2}{2}+\frac{\gamma}{\gamma-1}\frac{p}{\rho}=\frac{\gamma}{\gamma-1}\frac{p_\infty}{\rho_\infty}
\label{bern1}
\end{equation}
and the mass continuity equation
\begin{equation}
v=\frac{\dot M}{4\pi\rho r^2}, \label{nereq}
\end{equation}
where the constant $\dot M$ determines the rate of the BH mass increase and $\gamma$ is the polytropic index. The sound speed for the considered polytropic equation of state is
$c_s=\sqrt{\gamma p/\rho}$. In coordinates $(c_s,v)$, Eqs (\ref{bern1}) and (\ref{nereq})
respectively represent an ellipse and a hyperbola. They either do not intersect or intersect at one or two points. These curves always touch each other on the bisector at which the velocity of motion
becomes equal to the sound speed, and the subsonic flow becomes supersonic. We do not describe the Newtonian accretion properties in detail here because this case has been considered in much detail in many papers and textbooks (see, e.\,g., a very clear presentation in \cite{ZelNov73}.

\subsection{Relativistic accretion of a perfect fluid}
\label{Schwarzschild}

We now consider the relativistic accretion of a fluid with nonzero pressure. The Schwarzschild metric corresponding to a nonrotating noncharged BH with mass $M$ is given by
\begin{equation}
 \label{metrics}
 ds^2=fdt^2-f^{-1}dr^2-r^2(d\theta^2+\sin^2\!\theta\,d\phi^2),
\end{equation}
where
\begin{equation}
 \label{Shf}
 f=1-\frac{2M}{r}.\nonumber
\end{equation}
Below, we use units in which $c=G=1$. We consider the accretion of a perfect fluid with the energy-momentum tensor
\begin{equation}
 \label{emtensor}
    T_{\mu\nu}=(\rho+p)u_\mu u_\nu - pg_{\mu\nu},
\end{equation}
where $\rho$ and $p$ are the rest-frame density and pressure of the fluid and $u^\mu=dx^\mu/ds$ is the fluid 4-velocity normalized as $u^{\mu}u_{\mu}=1$. We assume that the pressure depends only on the density, $p=p(\rho)$, and temporarily consider this dependence arbitrary. Perfect fluid (\ref{emtensor}) describes a fairly wide class of matter exactly or to some approximation.

Michel \cite{Mic72} found a general relativistic solution for the spherically symmetric accretion of ordinary (baryonic) matter treated as a test fluid, ignoring back reaction on the metric, from the equations of mass and energy flux conservation in the Schwarzschild metric. The mass flux  conservation coincides with the particle number conservation in the case of a gas. When treating dark energy as a perfect fluid, no presence of particles is assumed at all. In this case, accretion calculations should be somewhat modified, such that no particle number conservation is required in general~\cite{BabDokEro04}. It is possible to formally introduce a function $n$, expressed through the dark energy equation of state as
$dn/n=d\rho/(\rho+p(\rho))$; if the medium consists of individual conserved particles, $n$
coincides with the particle number density~\cite{BabDokEro05-2}.

The projection of the energy-momentum tensor conservation law $T^{\mu\nu}_{\quad ;\nu}=0$ onto the 4-velocity direction, $u_{\mu}T^{\mu\nu}_{\quad ;\nu}=0$, yields the continuity equation for a perfect fluid:
\begin{equation}
 \label{eq2}
  u^{\mu}\rho_{,\mu}+(\rho+p)u^{\mu}_{\; ;\mu}=0.
\end{equation}
From (\ref{eq2}) we find the integral of motion (an analog of the energy conservation law)
\begin{equation}
 \label{fluxs2}
 ux^2 \frac{n}{n_\infty}=-A,
\end{equation}
where the dimensionless radius is $x\equiv r/M$,
\begin{equation}
\frac{n}{n_\infty}\equiv \exp\left[\;\,\int\limits_{\rho_{\infty}}^{\rho}\!\!
\frac{d\rho'}{\rho'+p(\rho')}\right], \label{ndef}
\end{equation}
$u=dr/ds<0$ in the case of direction to the center (accretion), and  $A>0$ is a dimensionless constant, to be found in Section~\ref{linsolsubsec}.

The integration of the time component of the conservation law $T^{0\nu}_{\;\;\; ;\nu}=0$ yields another integral of motion:
\begin{equation}
 \label{eq1}
 (\rho+p)(f+u^2)^{1/2}x^2u=C_1,
\end{equation}
where $u\equiv dr/ds$ and $C_1=const$. From (\ref{fluxs2}) and
(\ref{eq1}), it is straightforward to obtain
\begin{equation}
\frac{(\rho+p)}{n}(f+u^2)^{1/2}=C_2,
 \label{energy}
\end{equation}
where
\begin{equation}
\label{C2} C_2\equiv \frac{-C_1}{A}=\frac{\rho_\infty+p(\rho_\infty)}{n(\rho_\infty)},\nonumber
\end{equation}
and $\rho_\infty$ is the density at infinity. Equations (\ref{fluxs2}) and
(\ref{energy}) in combination with the equation of state of the fluid $p=p(\rho)$ make a closed systems of equations describing accretion of dark energy onto a BH.

During accretion, the BH mass changes as $\dot M=-4\pi r^2T_0^{\;r}$, which follows from the interpretation of the component $T_0^{\;r}$ in terms of the energy flux. Using (\ref{fluxs2})
and (\ref{energy}), the last equation can be transformed into
\begin{equation}
 \label{evol}
 \boxed{ \dot M=4\pi A M^2 [\rho_{\infty}+p_{\infty}]. }
\end{equation}
This is an important result, to be used in various sections of this review. Anticipating a discussion in what follows, we note that Eq (\ref{evol}) implies that the BH mass decreases during the accretion of phantom energy characterized by the condition $\rho_{\infty}+p(\rho_{\infty})<0$.

The constant $C_2$ is fixed by the boundary condition at infinity. From Eqs (\ref{fluxs2}), (\ref{energy}) and (\ref{C2}), we can find the density and velocity of the fluid at the event horizon. To calculate the constant $A$ in (\ref{fluxs2}) and, accordingly, the energy flux onto the BH, we make the physical assumption that the flow smoothly crosses the critical point (see \cite{Mic72,Bes97,Bes03,Bogovalov} for more details). Namely, by differentiating Eqs (\ref{fluxs2}) and (\ref{energy}), we obtain the relation
\begin{equation}
\frac{du}{u}\left[V^2-\frac{u^2}{1-\frac{2}{x}+u^2}\right]+\frac{dx}{x}\left[2V^2-\frac{1}{x\left(1-\frac{2}{x}+u^2\right)}\right]=0.
\label{crpoint}
\end{equation}
In the calculations, the parameter with the dimension of velocity,
\begin{equation}
\label{V} V^2=\frac{n}{\rho+p}\frac{d(\rho+p)}{dn} -1,
\end{equation}
first emerges, which, by virtue of (\ref{ndef}) is equal to the sound speed in the medium:
\begin{equation}
\label{V1} V^2=c_s^2(\rho)=\partial p/\partial\rho.
\end{equation}
For the solution to be single-valued, both square brackets in (\ref{crpoint}) must be equal to zero, and then the resulting equations determine the critical point. Thus, from (\ref{crpoint}), we find the critical point parameters:
\begin{equation}
\label{cpoint} u_*^2=\frac{1}{2 x_*},\quad V_*^2=\frac{u_*^2}{1-3u_*^2},
\end{equation}
where the subscript ``$*$'' marks values taken at the critical point.

\begin{figure}[t]
\begin{center}
\includegraphics[width=0.49\textwidth]{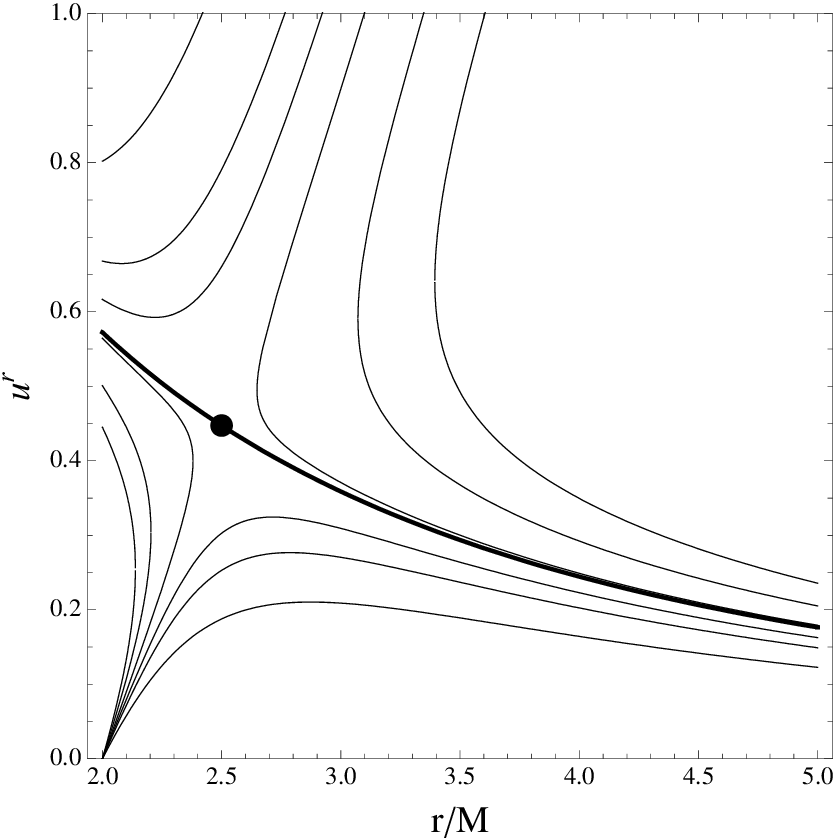}
\end{center}
\caption{\label{figcrit} Solutions for the radial velocity of a fluid with the equation of state $p=\rho/2$ as a function of radius in the case of accretion onto a Schwarzschild BH. Different curves correspond to different fluxes. The single-valued solution (bold curve) passes through the critical point marked by the black dot.}
\end{figure}

From Eqs (\ref{cpoint}), (\ref{V1}), (\ref{C2}) and (\ref{energy}) we find
\begin{equation}
 \label{rho_c}
 \frac{\rho_*+p(\rho_*)}{n(\rho_*)}=
 \left[1+3c_s^2(\rho_*)\right]^{1/2}
 \frac{\rho_\infty+p(\rho_\infty)}{n(\rho_\infty)},
\end{equation}
which yields $\rho_*$ for any $p=p(\rho)$. From the above equations, all other quantities of the problem, including the constant $A$ can be derived. We note that there is no critical point outside
the BH horizon ($x_*>1$). This fact has a simple interpretation: the solution has a critical point if the fluid velocity increases from subsonic to supersonic values. In the case $c_s^2<0$, the fluid velocity never crosses such a point. If $c_s^2>1$, the critical point can appear inside the event horizon (see Section~\ref{teprichsec}).

Fig.~\ref{figcrit} shows several solutions with different fluxes and the unique correct solution passing through the critical point.

\begin{figure}
\includegraphics[width=0.49\textwidth]{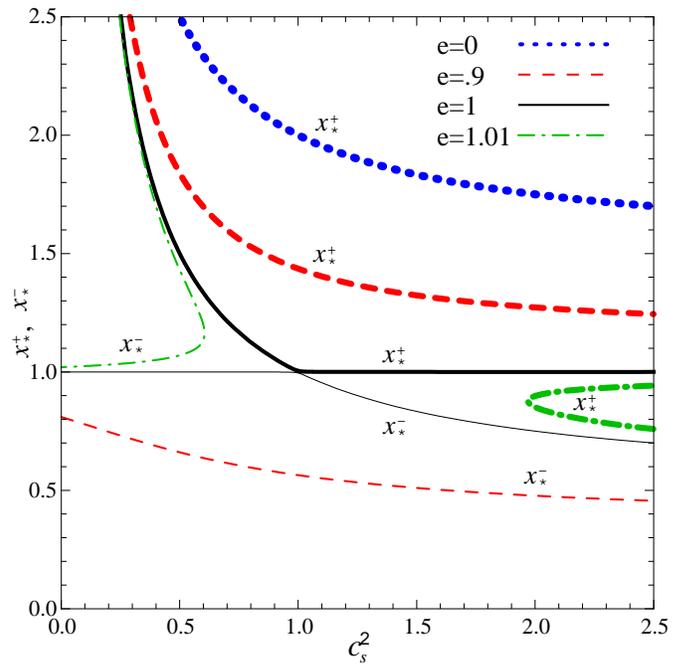}
\caption{The outer critical radius $x_*^+$ (bold lines) and the inner critical radius $x_*^-$ (thin lines) as functions of the sound speed $c_s$ for different values of the electric charge $e=Q/M$. The outer critical point coincides with the event horizon, $x_*^+=1$, 1, for an extreme BH ($e=1$) when $c_s\geq1$.}
 \label{Fig Rcr}
\end{figure}

The problem of accretion onto a BH considered here is self-consistent if:
 1) the accreting fluid is light, and 2) the BH mass increases slowly (the stationary limit). To satisfy these conditions, two parameters must be small. The first is the ratio of the mass of the fluid $M_{\rm gas} \sim \rho_\infty M^3$ inside the spherical volume with the BH gravitational radius to the BH
mass $M$: $\rho_\infty M^3/M=\rho_\infty M^2\ll1$. When this parameter is small, the test fluid approximation in the background metric is valid for the radii $r\ll R_{\rm max}=M(\rho_\infty M^2)^{-1/3}$. The second small parameter characterizes the slow rate of the BH mass change relative to the characteristic hydrodynamic time, $\dot M/M\ll c_s/M$, where $c_s$ is the sound speed in the accreting matter. According to (\ref{evol}), both parameters become equal, $\dot M\sim\rho_\infty M^2\ll1$, in the case of accretion of a perfect fluid, for which $|w-1|$ is not too close to zero and under the condition that $c_s$ is of the order of unity.

We now consider accretion onto a Reissner--Nordstr\"om BH with an electric charge $Q$. The Reissner-Nordstr\"om metric can be expressed in form (\ref{metrics}), but now with
\begin{equation}
 \label{RN1}
 f=1-\frac{2M}{r}+ \frac{Q^2}{r^2},\nonumber
\end{equation}
We set $e\equiv Q/M$. For $e^2<1$, the equation $f(x)=0$ has two roots:
\begin{equation}
 \label{hors}
 x_{\pm}=1\pm \sqrt{1 - e^2}.\nonumber
\end{equation}
The larger root $x=x_+$ corresponds to the Reissner--Nordstr\"om BH event horizon, and $x=x_-$ is the so-called Cauchy horizon, or the inner horizon. In the opposite case, $e^2>1$ metric (\ref{metrics}) describes the so-called naked singularity, which is not hidden by the event horizon from an external observer. The degenerate case $e^2=1$ corresponds to an extreme BH.

Using the above method, we obtain the relations at the critical point:
\begin{equation}
 \label{cpoint*}
  u_*^2=\frac{x_*-e^2}{2x_*^2}, \quad
  c_s^2(\rho_*)=\frac{x_*-e^2}{2x_*^2-3x_*+e^2}.
\end{equation}
From (\ref{cpoint*}) we obtain
\begin{equation}
 \label{xc}
 x_*^{\pm} =  \frac{1+3c_*^2}{4c_*^2}
 \left\{1\pm\left[1-\frac{8c_*^2(1+c_*^2)}{(1
 +3c_*^2)^2}e^2\right]^{1/2}\right\},
\end{equation}
where $c_*\equiv c_{s}(x_*)$. Critical points exist, if
\begin{equation}
 \label{q}
e^2\le \frac{\left(1+3c_*^2\right)^2}{8c_*^2\left(1+c_*^2\right)}.\nonumber
\end{equation}
We note that here, in contrast to a unique critical point in the Schwarzschild BH case, there are formally two critical points corresponding to the plus and minus signs in (\ref{xc}), and $x_{*}^{-}\to0$ as $e\to 0$. Depending on the values $e$ and $c_s$, five distinct cases of the mutual location of the horizon and critical points can be realized \cite{BabDokEro11}.

In Fig.~\ref{Fig Rcr}, the critical radius is shown as a function of the sound speed for different values of $e$.


\subsection{Accretion of a fluid with a linear equation of state}
\label{linsolsubsec}

The equation of state for dark energy modelled as a perfect fluid is frequently written in the form
$p=w\rho$, where $w=const<0$. In this case, however, the medium is dynamically unstable, because the square of its sound speed is negative: $c_s^2=\partial p/\partial\rho=w$. In field models, the equation of state and the square of the sound speed are not related in such a way in general. However, it is much more convenient to solve accretion problems using the perfect fluid approximation, and it is therefore desirable to circumvent this difficulty. For this, it is convenient to start with a perfect fluid with a more general equation of state
\begin{equation}
  \label{pDE}
  p=\alpha(\rho-\rho_0),
\end{equation}
where $\alpha$ and $\rho_0$ are parameters. This simple generalization of the linear equation of state allows considering dark energy, including phantom energy, with a positive square of the sound speed, $c_s^2=\alpha\geq0$, although $w$ can here be negative. Interestingly, for $\alpha=-1/3$ Whittaker found an exact static solution --- a stable spherically symmetric field configuration  \cite{Whi68}. The value $\alpha=-1/3$ is special because in this case the combination $\rho+3p=\rho_0$, which serves as a source in gravitational field equations, takes a constant
value.

The evolution of a universe filled with dark energy with equation of state (\ref{pDE}) was studied in \cite{BabDokEro05-1}. Assuming this equation of state to be valid, interesting solutions can be
derived, including an anti-Big Rip or bounce (the change of contraction with expansion).

Equation of state (\ref{pDE}) can be reduced to an `effective' cosmological constant and the dynamically evolving dark energy by redefining the density and pressure as
\begin{eqnarray}
  \label{repl}
  \rho=\rho_\Lambda+\rho_\alpha, \quad
  p=p_\Lambda+p_\alpha,
\end{eqnarray}
where $p_\Lambda=-\rho_\Lambda$, $p_\alpha=\alpha\rho_\alpha$ and
\begin{equation}
  \label{rhoLN1}
  \rho_\Lambda=\frac{\alpha\rho_0}{1+\alpha}, \quad
  \rho_\alpha=\rho-\frac{\alpha\rho_0}{1+\alpha}.
\end{equation}
We note that such a separation into two effective fluids can be done only for a strictly linear equation of state.
\begin{figure}[t]
\begin{tabular}{c c}
\includegraphics[angle=0,width=0.24\textwidth]{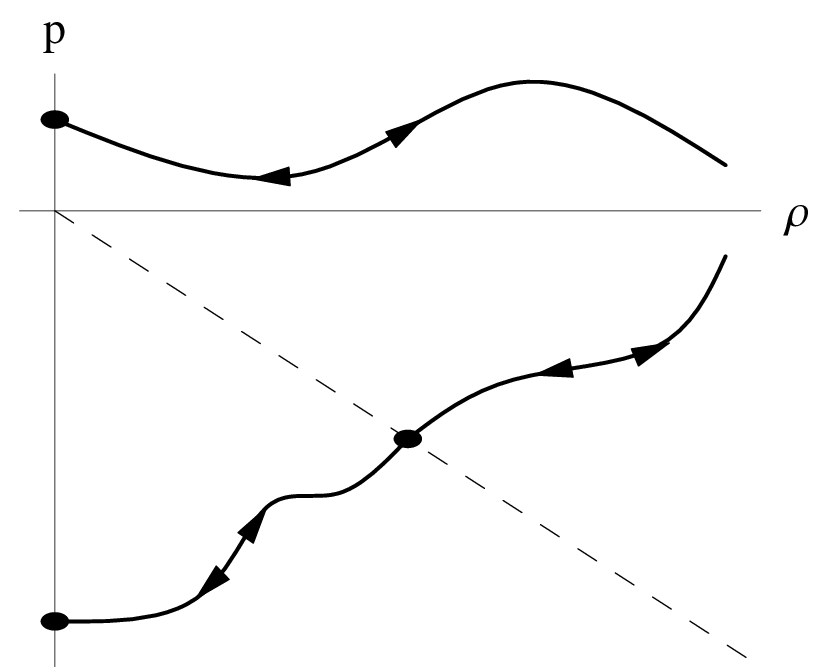}&
\includegraphics[angle=0,width=0.24\textwidth]{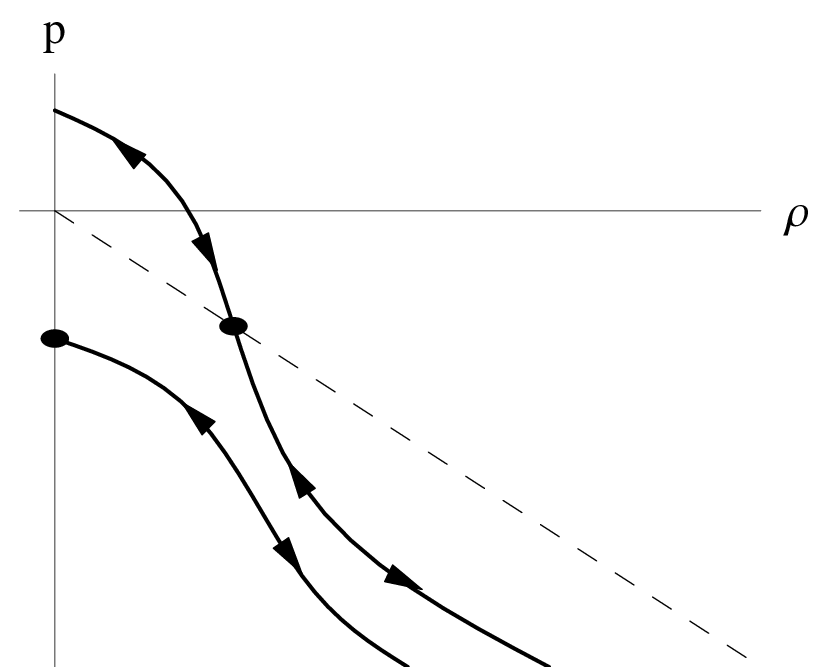}\\
\end{tabular}
  \caption{\label{Gen1}
Sketch of the evolution of dark energy with an arbitrary equation of state $p(\rho)$. At each point, the curve $p=p(\rho)$ can be approximated by linear dependence (\ref{pDE}).}
\end{figure}

Instead of (\ref{pDE}), we can consider an arbitrary smooth curve $p=p(\rho)$, shown in Fig.~\ref{Gen1}. Because any smooth curve in the vicinity of its point can be approximated by a linear function, Eq (\ref{pDE}) can be considered a linear approximation of the general nonlinear equation of state $p=p(\rho)$ near some point $\rho=\rho_1$, as long as $|\rho-\rho_1|$ is sufficiently small. In particular, if the curve $p=p(\rho)$ intersects the $\Lambda$-term line, a universe with dark energy always approaches the de Sitter attractor state \cite{BabDokEro05-1}.
\begin{widetext}
\begin{center}
\begin{figure}
\includegraphics[angle=-90,width=0.95\textwidth]{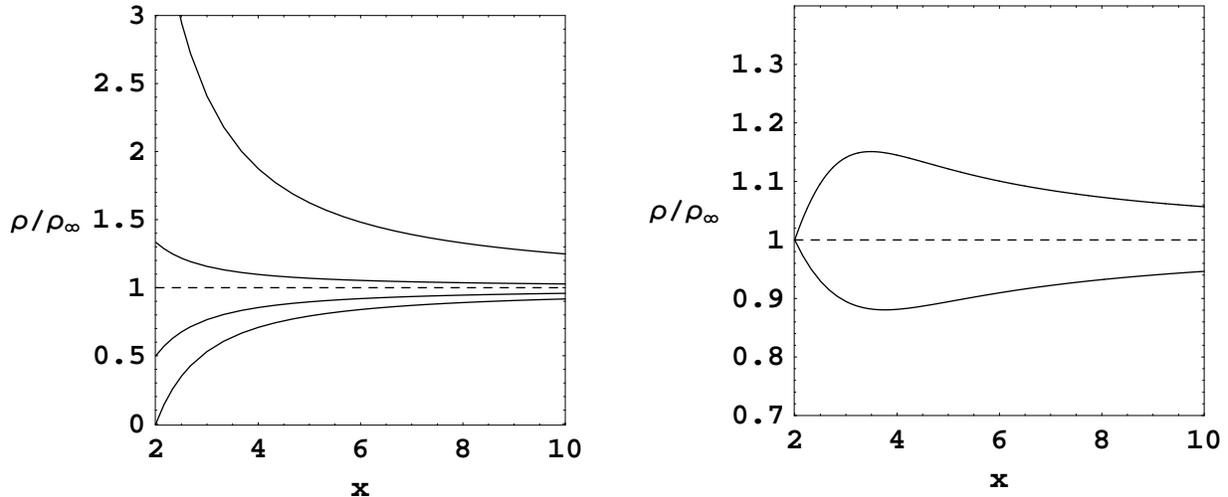}
  \caption{\label{Figrhol}
The density (normalized to the value at infinity) $\rho/\rho_{\infty}$, as a function of the coordinate $x$ for linear model (\ref{pDE}). (a) The density $\rho/\rho_\infty$ of a
hydrodynamically stable fluid with $\alpha=1$1: curve 1, $\rho_0=0$ (model of neutron star matter); curve 2, $\rho_0/\rho_\infty=16/9$ (linear model of non-phantom dark energy); curve 3, $\rho_0/\rho_\infty=7/3$ (linear model of phantom energy); curve 4, $\rho_0/\rho_\infty=7/3$ (linear model of phantom energy with the density at the horizon $\rho_{\rm H}=0$). The density $(\rho/\rho_\infty)$ at $\alpha<0$: curve 1, $\alpha=-2$, $\rho_0=0$ and $A=4$ (linear model of phantom energy); curve 2, at $\alpha=-1/2$, $\rho_0=0$ and $A=4$ (linear model of non-phantom energy).}
\end{figure}
\end{center}
\end{widetext}

We consider the accretion of a perfect fluid with equation of state (\ref{pDE}) onto a Reissner--Nordstr\"om BH. From (\ref{fluxs2}), using (\ref{cpoint}) and (\ref{xc}), we can find the dimensionless constant $A$ for the linear equation of state,
\begin{equation}
 A=\alpha^{1/2}x_*^2 \left(\frac{2\alpha x_*^2}{x_*-e^2} \right)^{{\frac{\scriptstyle 1-\alpha}{\scriptstyle 2\alpha}}}.
 \label{Alinear}
\end{equation}
The velocity and energy density as a function of radius are determined from solutions (\ref{flux}) and (\ref{energy}) using the following relations
\begin{equation}
 f+u^2=\!\left(\!-\frac{ux^2}{A}\right)^{\!2\alpha}\!\!, \quad
  \frac{\rho\!+\!p}{\rho_\infty+p_\infty}=\!
  \left(\!-\frac{A}{ux^2}\right)^{\!1+\alpha}\!.
  \label{urholinear}
\end{equation}
Solutions of these equations can be expressed in terms of analytic functions in special cases where $\alpha=1/4$, $1/3$, $1/2$, $2/3$, $1$, $3/2$ and $2$. For example, for $\alpha=1/3$ (thermalized photon gas),
\begin{equation}
 \rho=\frac{\rho_0}{4}+\left(\rho_{\infty}-\frac{\rho_0}{4}\right)  \left(\frac{1+2z}{3f}\right)^2,\nonumber
 \label{alpha13}
\end{equation}
where
\begin{equation}
 z=\left\{\
\begin{array}{lr}
 \displaystyle{\cos\frac{2\pi-\beta}{3}\,}, &
 x_+\leq x\leq x_*,\\ \\
 \displaystyle{\cos\frac{\beta}{3}}, & x>x_*,
 \label{rho}
\nonumber
\end{array}
 \right.
\end{equation}
\begin{equation}
 \beta=\arccos\left(1-\frac{27}{2}A^2\frac{f^{\,2}}{x^4}\right).\nonumber
 \label{beta}
\end{equation}
The obtained expressions are simplified for a Schwarzschild BH. For example, the critical point parameters in this case are $x_*=(1+3\alpha)/(2\alpha)$, $u_*^2=\alpha/(1+3\alpha)$. The constant that determines the flux onto the BH takes the form
\begin{equation}
\label{A1} A=\frac{(1+3\alpha)^{(1+3\alpha)/2\alpha}}{4\alpha^{3/2}}.
\end{equation}
It is easy to see that $A\ge 4$ for $0<\alpha<1$, while for $\alpha>1$ it can be shown that $A<4$. At $\alpha=1$, we have $A=4$. These considerations lead to the conclusion that for typical sound
velocities, the constant $A$ is of the order of unity. Fig.~\ref{Figrhol} shows the fluid density as a function of $x$ \cite{BabDokEro05-2}.

\begin{figure}
\includegraphics[width=0.49\textwidth]{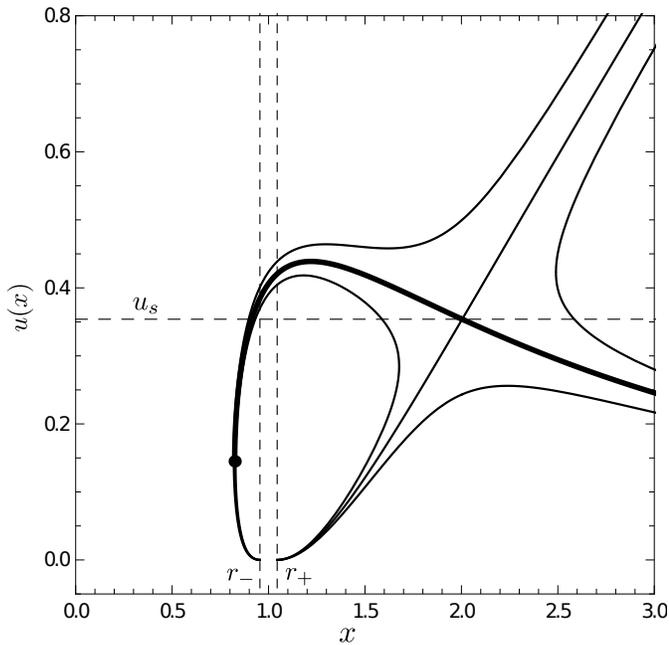}
\caption{The radial 4-velocity $u(r)$ (bold curve) for an accreting fluid with $p/\rho=1/3$ (thermalized photon gas) in the Reissner--Nordstr\"om metric with the charge $e=0.999$, $u_s$ is the 4-velocity at the critical point.}
 \label{rmine}
\end{figure}

The case of a superluminal fluid is of interest for a Reissner--Nordstr\"om BH. `Superluminal' dark energy is discussed in more detail in Section~\ref{teprichsec}. Here, we only mention
that the behavior of the superluminal fluid ($c_s>1$) is rather unusual. There is an infinite family of regular solutions at $r>0$, which are parameterized by the constant $A$. Each solution includes one hydrodynamic branch, and there is no sound horizon.

The solution for a subsonic fluid exists only inside the region with some minimal radius $r>r_{\rm min}$, where  $0<r_{\rm min}<r_-$; the accreting fluid does not reach the central singularity, and its density reaches a maximum at $r_{\rm min}$, shown in Fig.~\ref{rmine}. A similar behavior was discovered for test particles with a nonzero mass moving along geodesics \cite{carter68,Lopez} in the Reissner--Nordstr\"om metric, with the particles bouncing at the radius $r_{\rm min}=Q^2/(2M)$. The corresponding solutions for accretion of a subluminal fluid are singular at $r=r_{\rm min}$, namely, $u'(r_{\rm min})=\infty$ and $\rho'(r_{\rm min})=-\infty$ (although the 4-velocity and the density remain finite at $r=r_{\rm min}$). As a result, continuity equation (\ref{eq2}) is ill-defined at $r=r_{\rm min}$.

\section{Accretion of phantom energy and the fate of black holes}
\label{cosmsec}

In Section~\ref{linsolsubsec}, we showed that during the accretion of a perfect fluid with $\rho+p>0$ the masses of BHs increase as in the case of accretion of ordinary matter. But a qualitatively different result follows for phantom energy --- a medium with $\rho+p<0$. Equation (\ref{evol}) implies that the BH mass decreases in this case. In this Section~\ref{cosmsec} we discuss the notion of phantom energy and its properties, and then in Section~\ref{cosmsec32} we consider accretion of phantom energy onto BHs.


\subsection{Violation of energy conditions and phantom energy}
\label{cosmsec31}

Before the appearance of the notion of phantom energy, it was usually assumed in studying the general properties of solutions of the Einstein equations that the energy conditions hold \cite{hawkell}, which were thought to be appropriate for physically admissible matter. These conditions underlie general theorems about singularities and horizons. However, more exotic cases where the energy conditions are violated have recently been discussed in numerous papers. Even if the real cosmological dark energy is not the phantom energy, studying the models with violated energy conditions is interesting from the theoretical point of view and turns out to lead to nontrivial results.

The weak energy condition can be formulated as follows: for any timelike vector $v^a$ the inequality $T_{ab}v^av^b\ge 0$ holds. For a perfect fluid with energy-momentum tensor (\ref{emtensor}), this
implies that $\rho\ge 0$ и $\rho+p\ge 0$. As was proved by Christodoulou \cite{1970} (1970) and Hawking \cite{1971} (1971), when the weak energy condition is satisfied, the surface area of a
BH does not decrease for any classical (nonquantum) processes. For the phantom energy, the weak energy condition is violated, and this theorem is invalid. Therefore, the BH mass can decrease during accretion of phantom energy.

Different field theory models have been proposed for the phantom energy. In theories with a scalar field, the corresponding Lagrangian must have a negative kinetic term \cite{Caldw1,Caldw2}, for example, $K=-\frac12\phi_{;\mu}\phi^{;\mu}$. Then the phantom energy flux onto a BH has the opposite sign, $T_{0r}=-\phi_{,t}\phi_{,r}$, where $\phi$ is the solution of the same Klein--Gordon equation as in the case of the standard scalar field. In this case, accretion of a scalar phantom field causes the BH mass to decrease at the rate $\dot M= -4\pi(2M)^2{\dot{\phi}}^2_\infty$. A more general form of the negative kinetic term in the $k$-essence model was considered in \cite{GonK04}.

The simplest models of phantom energy are unstable at the quantum level due to the appearance of ghost solutions. In more sophisticated models with (dynamical) Lorentz symmetry violation, it is possible to avoid the catastrophic quantum instability of phantom energy~\cite{Rub06,Libetal07}. In these models, the instability of phantom fields occurs only at low energies \cite{Rub06,Libetal07}. Moreover, in models with the Galileon, the phantom regime emerges quite naturally without the appearance of ghosts or a gradient instability \cite{Deffayet:2010qz}.
As a result, physically acceptable models of phantom energy have been constructed.

We also note that in addition to field phantom models, phantom energy can be mimicked by deviations from GR in $f(R)$-gravity and scalar-tensor models of gravity. Studies of phantom energy have their own general theoretical meaning for analysing possible properties and paradoxes of phantom energy, as well as for clarifying physical features and conditions under which the `pathological' behavior appears.

\subsection{Accretion of phantom energy}
\label{cosmsec32}

Formula (\ref{evol}) implies that the BH mass must decrease due to the accretion of phantom energy. This result is independent of the equation of state $p=p(\rho)$; only the condition $p+\rho<0$, is
important, under which the phantom energy falling into the BH brings energy outwards. We recall that $\rho$ and $p$ in the rest-frame of the fluid, i.\,e., as measured by the comoving observer, enters the energy-momentum tensor $T_{\mu\nu}=(\rho+p)u_\mu u_\nu - pg_{\mu\nu}$. But if the fluid moves with a velocity $v$, relative to the observer, the observer measures the energy density \cite{DokEro10} (see also \cite{Sawicki:2012pz}, where similar considerations are applied to Galileons):
\begin{equation}
\rho'=T_{0'0'}=\frac{\rho+pv^2}{1-v^2}, \label{tmoove}
\end{equation}
in accordance with the usual Lorentz transformations (see \cite{LL}, \S~35). Bearing this in mind, it is easy to understand that the observer at rest in the Schwarzschild metric near the BH horizon (where the physical velocity of the fluid $v\to 1$) in the phantom case $\rho+p<0$ measures the flux onto the BH with a negative energy density, $\rho'<0$. This example clearly explains the reason for the decrease in the BH mass.

If we neglect the cosmological evolution of  $\rho_\infty$, then we obtain from (\ref{evol}) that
\begin{equation}
 \label{m}
 M=M_i\left(1+\frac{t}{\tau}\right)^{-1},
\end{equation}
where $M_i$ is the initial mass of the black hole and the characteristic time of evolution is
\begin{equation}
 \label{tau} \tau=\frac{-1}{4\pi A M_i [\rho_\infty+p(\rho_\infty)]}.
\end{equation}
Equation (\ref{m}) implies that for $\rho_\infty+p(\rho_\infty)<0$ the BH mass decreases with time.

\subsection{Thermodynamics of phantom energy}
\label{phtersubsec}

In a medium with $p\le0$, such as dark energy, some nontrivial energetic effects can occur \cite{Iva09}. For example, during an adiabatic expansion, the internal energy increases, in contrast to ordinary matter with $p\ge0$. Phantom energy, if it exists, has even more exotic thermodynamic properties. We consider the general thermodynamic relation for dark energy:
\begin{equation}
\label{tmain} TdS=pdV+dE,
\end{equation}
where $V=\int\sqrt{\gamma}d^3x$, $\gamma_{\alpha\beta}$ is the spatial metric tensor and the
integration is performed over the volume with a size smaller than the characteristic scale of change of the relevant quantities. For a small comoving volume, Eq (\ref{eq2}) implies that $nV=const$. It is useful to introduce specific values $\rho=E/V$ and $\sigma=S/(nV)$. Under the adiabatic condition $d\sigma=0$, which is equivalent to the definition $d\rho=(\rho+p)dn/n$ introduced in (\ref{tmain}), it is possible to find the following relation from (\ref{tmain}), which is a particular case of the general relation found in \cite{SilLimCal02} for a dissipative medium:
\begin{equation}
\label{dtt} \frac{dT}{T}=\frac{dn}{n}\left(\frac{\partial p}{\partial\rho}\right)_n.
\end{equation}
For the linear equation of state (\ref{pDE}) it follows from (\ref{dtt}) that
\begin{equation}
n=const\cdot|T|^{1/\alpha}\label{n}
\end{equation}
Setting $\rho_{\Lambda}=\alpha\rho_0/(1+\alpha)$, , it is straightforward to derive
from (\ref{n}) that
\begin{equation}
\rho=\rho_{\Lambda}\pm C_3|T|^{\frac{1+\alpha}{\alpha}},
\end{equation}
with a constant $C_3>0$, the plus sign corresponding to $\rho+p>0$, and the minus sign corresponding to $\rho+p<0$.

We now find the expression for the entropy of phantom energy. If the chemical potential $\mu$ satisfies the equality $d\mu=-\sigma dT+Vdp/N=0$, then
\begin{equation}
\sigma=\frac{1}{n}\frac{d p}{d T}. \label{sigeq}
\end{equation}
In the adiabatic case $\sigma=const$, using the relation $(\partial p/\partial
T)_\sigma=(\partial p/\partial \rho)_\sigma(\partial\rho/\partial T)_\sigma$, it finally follows from (\ref{sigeq}) that
\begin{equation}
S=\pm V(1+\alpha)C_3|T|^{\frac{1+\alpha}{\alpha}}T^{-1}.
\end{equation}
Hence, the entropy of phantom energy is positive, $S>0$ while the temperature is negative, $T<0$  (this case was first considered in \cite{GonSig04}), and vice versa.

Negative temperatures are considered in physics, for example, in application to the inverse population of quantum levels, when the number of particles at higher levels is larger than that at lower levels. In practice, this situation occurs for some subsystems of electrons in the working matter of a laser. Similarly, negative temperatures can be realized for the degrees of freedom of translation motion of atoms in an ultracold gas \cite{Braetal13}.

In classical physics, the entropy defined in terms of the statistical weight is nonnegative. However, the notion of negative entropy arises in quantum mechanics in systems with quantum entanglement \cite{CerAda97}. Therefore, phantom energy can reflect some specific quantum properties of its physical constituent.

We consider the entropy balance during accretion using the well-known relations for the temperature and entropy of a BH:
\begin{equation}
T_{\rm BH}=\frac{\hbar c^3}{8\pi GM},~~~~~S_{\rm BH}=\frac{4\pi GM^2}{\hbar c}.
\end{equation}
We consider the sphere of a radius $r^*\gg 2M$ around a Schwarzschild BH. We mark the quantities related to the interior and exterior of this sphere by the subscripts `in' and `out'. If the mass of dark energy can be neglected (as we have assumed so far), then $\dot S_{\rm in}\ll\dot
S_{\rm out}$. For the external region,
\begin{equation}
\dot S_{\rm out}\simeq 4\pi r^{*2}\sigma u=-4\pi M^2A\sigma_{\infty}=-\frac{\dot M}{T},
\end{equation}
where we use the stationarity of the flux onto the BH. The total entropy is
\begin{equation}
\dot S=\dot S_{\rm BH}+\dot S_{\rm in}+\dot S_{\rm out}\approx\dot M\left(\frac{1}{T_{\rm
BH}}-\frac{1}{T}\right)\label{sacitog}
\end{equation}
Hence, the total entropy does not change during accretion, $\dot S=0$, up to a small value $\dot
S_{\rm in}$, if $T_{\rm BH}=T$, i.\,e., if some kind of thermodynamic equilibrium between dark energy and the BH is established.

It is interesting to consider the formal problem of dark energy enclosed in a cell with impermeable walls \cite{Braetal13}. If the energy of the system is conserved, then, from (\ref{tmain}) and the
maximum entropy principle, we obtain
\begin{equation}
\frac{p}{T}=\left.\frac{\partial S}{\partial V}\right|_E\ge0.
\end{equation}
This implies that the signs of pressure and temperature are the same and a medium with $p<0$ must have a negative temperature, $T<0$.

Different aspects of the dark energy thermodynamics have also been discussed in other papers (see, e.\,g., \cite{LimAlc04,Siletal12,GonSig04,IzqPav06,PerLim08,LimPer08,BreNojOdiVan}.

\subsection{Fate of black holes in a universe approaching the Big Rip}

We now discuss the evolution of black holes in a universe with the Big Rip when the scale factor
$a(t)$ increases to infinity in a finite time interval \cite{Caldw1,Caldw2}. We consider the epoch in which only dark energy is important and other forms of matter can be ignored. Setting  $\rho_0=0$ for simplicity in linear model (\ref{pDE}), we find the law of phantom density evolution in such a
universe:
\begin{equation}
\label{sol3} \rho_\infty=\rho_{\infty,i}\left(1-\frac{t}{\tau}\right)^{-2},
\end{equation}
where
\begin{equation}
\label{tau2} \tau^{-1}=-\frac{3(1+\alpha)}{2}\left(\frac{8\pi}{3}\rho_{\infty,i}\right)^{1/2}
\end{equation}
$\rho_{\infty,i}$ is the initial density of the cosmological phantom energy, and the initial time is chosen such that the Big Rip occurs at the time $\tau$. In particular, a rapid increase in the phantom energy density and the value of the scale factor can falsify the current astronomical prediction that our Galaxy will collide with the Andromeda nebula in a few billion years. By contrast, starting from some time, the galaxies will recede with acceleration.

We note that a single condition $\rho+p<0$ is insufficient for the evolution to end up with a Big Rip \cite{McInnes1,McInnes2}. Examples of phantom cosmology without the Big Rip are considered, for example, in \cite{McInnes1}.

From Eq (\ref{evol}) and using (\ref{sol3}), we derive the evolution of the BH mass in a universe approaching the Big Rip:
\begin{equation}
 \label{mevol1}
 M=M_i\left(1+\frac{M_i}{\dot M_0 \;\tau}\;
 \frac{t}{\tau-t}\right)^{-1},
\end{equation}
where
\begin{equation}
\dot M_0=(3/2)\,A^{-1}|1+\alpha|,
\end{equation}
and $M_i$ is the initial BH mass. For example, for $\alpha=-2$ and the typical value $A=4$
(corresponding to $u_{\rm H}=-1$), we have $\dot M_0=3/8$. In the limit $t\to\tau$ (i.\,e., near the Big Rip), the dependence of the BH mass on time $t$ becomes linear, $M\simeq\dot
M_0\,(\tau-t)$. As $t$ approaches $\tau$ the rate of the BH mass decrease is no longer dependent on its initial mass and the phantom energy density: $\dot M\simeq-\dot M_0$. In other words, the
masses of all BHs close to the Big Rip become almost equal. This means that the accretion of phantom energy dominates over the Hawking evaporation until the BH mass reduces to the Planck value. However, formally all BHs will be evaporated via the Hawking radiation at the Planck time before the Big Rip. Such is the fate of BHs in a universe approaching the Big Rip. Unlike all other objects, including elementary particles, which will be disrupted before the Big Rip, BHs, according to classical theory, must disappear (their masses will vanish) exactly at the instant of the Big Rip, and with the Hawking evaporation taken into account, by the Planck time interval before the Big Rip.

Similar problems of the fate of a BH at the time of bounce in the model of a pulsating universe and about possible observational manifestations of the `surviving' BHs in the next phase of the cosmological expansion are discussed in the literature. These include, for example, the traces of
collisions of BHs in the form of concentric circles appearing in the CMB temperature distribution \cite{GurPen11} (we note, however, that the circles found in \cite{GurPen11} have not been confirmed by other independent analyses). If the masses of BHs exceed some critical value immediately before the bounce, they can merge into one big BH \cite{CarCol11}. Accretion of dark energy can significantly change the BH evolution at the bounce points. Accretion of a non-phantom dark energy with $\rho+p>0$ at the stage of compression could overcome the BH mass decrease due to the Hawking evaporation, which would lead to the `survival' of low-mass BHs during the bounce, but they will also be evaporated at the next stage of expansion, when the accretion rate decreases.

\section{Noncanonical scalar fields and black holes}
\label{teprichsec}

In this section, we discuss the behavior of different noncanonical scalar fields in the vicinity of BHs. The study of noncanonical scalar fields is basically motivated by the dark energy problem: such fields have usually been proposed in the context of dark energy. Here, we are primarily interested in the behavior of such a field near a BH. To tackle this problem, we partially borrow the formalism of calculations of matter accretion onto BHs described in Sections~I--III. However, we are interested not in the accretion itself but in the way of finding scalar field solutions in the BH metric. First, we show how accretion of a perfect fluid can be represented in terms of the accretion of a $k$-essence scalar field, and then we study physical effects emerging when noncanonical scalar fields are present near a BH.

\subsection{Perfect fluid as a scalar field}
\label{perfect}

It is well known that potential flows of a relativistic perfect fluid can be described in terms of a scalar field \cite{Luk80}. In particular, an ultrarelativistic fluid corresponds to the canonical massless scalar field. To represent more complicated equations of state, it is necessary to introduce a scalar field with a more complicated noncanonical Lagrangian, written as
\begin{equation}
 \label{sfaction}
 \mathcal{L}=\mathcal{L}(X),\quad \mbox{где} \quad
 X\equiv\frac{1}{2} \partial_\mu\phi\partial^\mu\phi.
\end{equation}
The energy-momentum tensor corresponding to Lagrangian (\ref{sfaction}) has the form
\begin{equation}
T_{\mu\nu}=\mathcal{L}_{,X}\nabla_\mu\phi\nabla_\nu\phi - g_{\mu\nu}\mathcal{L}, \nonumber
\end{equation}
where the subscript $X$ denotes the derivative with respect to $X$. The correspondence between the scalar field and the perfect fluid with energy-momentum tensor (\ref{emtensor}) can be obtained by the identification (see, e.\,g., \cite{BabMukVik08}):
\begin{equation}
 u_{\mu}\equiv\frac{\nabla_{\mu}\phi}{\sqrt{2X}}.
 \label{unublapsi}
\end{equation}
The pressure $p$ coincides with the density of the scalar field Lagrangian
\begin{equation}
p=\mathcal{L}(X),  \label{p}
\end{equation}
and the density is written as
\begin{equation}
\rho\left(X\right)=2X\mathcal{L}_{,X}-\mathcal{L}.\label{e}
\end{equation}
The sound speed is then expressed as
\begin{equation}
 c_s^2=\frac{\mathcal{L}_{,X}}{\rho_{,X}}= \left(1+2X  \frac{\mathcal{L}_{,XX}}{\mathcal{L}_{,X}} \right)^{-1}.
 \label{sound in kessence}
\end{equation}
Besides the density $\rho$ and pressure $p$ the `particle number density' and enthalpy can be formally defined as
\begin{equation}
 n\equiv\exp\left(\int\frac{d\rho}{\rho +p}\right)=\sqrt{X}\mathcal{L}_{,X}.\nonumber
\end{equation}
and enthalpy
\begin{equation}
 h\equiv\frac{\rho+p}{n}=\frac{d\rho}{dn}=2\sqrt{X}.\nonumber
\end{equation}
Lagrangian (\ref{sfaction}) yields the equations of motion
\begin{equation}
\label{sfeom}
 \partial_\mu\left( \sqrt{-g}\, \mathcal{L}_X
\,g^{\mu\nu}\partial_\nu\phi \right)=0,
\end{equation}
The stationary flux can be determined by the ansatz
\begin{equation}
\phi(t,r)=\dot\phi_\infty t + \psi(r), \label{ansatzinf}
\end{equation}
where the constant $\dot\phi_\infty$ determines the `cosmological' value $\dot\phi$ at the spatial infinity. It is easy to verify that for ansatz (\ref{ansatzinf}) the following equation holds:
\begin{equation}
\label{X} X=\frac{1}{2}\left(\frac{\dot\phi_\infty^2}{f} - f\psi'^2\right),\nonumber
\end{equation}
and equation of motion (46) can be integrated. As a result, we obtain
\begin{equation}
\label{sfeom1}
  r^2 f \mathcal{L}_X  \psi'(r)= r_g^2 \dot\phi_\infty \tilde{A},
\end{equation}
where the coefficients in the right-hand side are chosen such that the parameter $\tilde A$, responsible for the energy flux, is dimensionless. Equation (\ref{sfeom1}) is an analog of Eq (\ref{fluxs2}), written in terms of a scalar field. Moreover, Eq (\ref{sfeom1}) is an algebraic equation for $\psi'$ (after expressing $\mathcal{L}_X$ in terms of $\psi'$). Therefore, the general solution contains the parameter $\tilde A$, which must be determined in a way similar to the determination of critical point parameters (\ref{cpoint}). From (\ref{sfeom}), we express $\psi''$ in terms of $\psi'$ (this expression also contains $\L_X$ and $\L_{XX}$). The critical point can be found by equating both the numerator and denominator of the obtained expression to zero. As a result, we find
\begin{equation}
\label{cpoint1}
 \psi_*'^2=\dot\phi_\infty^2 \frac{ r_*f_*'}{f^2\left(r_* f_*'+4 f_*\right)},
 \quad f_*\psi_*'^2 \mathcal{L}_{XX}=\mathcal{L}_X.
\end{equation}
which is similar to Eq (\ref{cpoint}). Hence, we have three equations (\ref{sfeom1}),
(\ref{cpoint1}), from which $\psi_*'$, $r_*$ and $A$ can be found. This procedure is fully equivalent to fixing the critical point of the accreting fluid. The accreting scalar field flux can be found as $\dot M = 4\pi r^2 T^r_t $, and therefore we ultimately obtain
\begin{equation}
\label{evol1}
\dot M = 4\pi \dot\phi_\infty^2 r_g^2 \tilde{A}.
\end{equation}
which coincides with Eq (\ref{evol}), up to a redefinition of $\tilde{A}$.

Let us consider Eq (\ref{sfeom1}) in the limit $x\equiv r/M\to 0$. We have $ 2X\sim  x^2B^2/e^2 -e^2\psi'^2/x^2$, where $B=const$. For the fluid, we have $X>0$, whence $X\to 0$ and $\psi'^2\to 0$ as $x\to 0$. On the other hand, we find from (\ref{sfeom1}) that $\mathcal{L}_X\psi'\to{\rm const}$ as $x\to 0$. By combining these equations, we arrive at the conclusion that the fluid reaches the coordinate $x=0$ during a stationary accretion process only if $\mathcal{L}_X\to \infty$ as $X\to0$. It hence follows, in particular, that a fluid described by a linear equation of state with $\alpha\leq 1$, never reaches the central singularity $x=0$, in the case of a stationary accretion process if $e\neq0$.

Wave fields of various types in the gravitational field of a BH have been considered in many papers. Especially well elaborated is the scattering of fields in the BH gravitational field, including in the form of `superradiation' --- field enhancement due to the BH rotational energy. We are
interested in the particular case of the behavior of fields near a BH when the energy flux through the horizon (accretion) is present. The scalar field Lagrangian is $L=K-V$, where $K$ is
the kinetic term and $V$ is a potential. For the standard form of the kinetic term $K=\frac12\phi_{;\mu}\phi^{;\mu}$ the energy flux is $T_{0r}=\phi_{,t}\phi_{,r}$. In \cite{Jac} the solution for the Schwarzschild metric was found for the zero potential $V=0$: $\phi= \dot\phi_\infty[t+2M\ln(1-2M/r)]$, where $\phi_\infty$ is the scalar field at infinity. In \cite{FroKof}, this solution was shown to be approximately valid for some scalar fields with a nonzero potential $V(\phi)$. In calculating the accretion of a scalar field with the canonical
kinetic term, we have $T_0^{\;r}=-(2M)^2\dot\phi^2_\infty/r^2$, and, accordingly, $\dot
M=4\pi(2M)^2{\dot{\phi}}^2_\infty$. The energy-momentum tensor constructed using the solution in \cite{Jac}, exactly coincides with that of a perfect fluid with the ultrahard equation of state $p=\rho$ after the substitution $p_{\infty}\to\dot{\phi}_\infty^2/2$, $\rho_{\infty}\to\dot{\phi}_\infty^2/2$.

\subsection{Induced metric and causal structure}
\label{induced}

Theories with a nontrivial kinetic structure allow the propagation of perturbations on the background of a nontrivial solution with a velocity different from the speed of light. In particular, superluminal propagation is possible. But despite the presence of superluminal signals, no causal paradoxes arise in these theories \cite{BabMukVik08,BabMukVik06,Bruneton:2006gf,Bruneton:2007si}.

In recent years, spontaneous breaking of the Lorentz invariance and related topics were attracting much interest. One of the main questions is whether the theories with superluminal propagation are self-consistent and whether they respect the causality principle, for example, due to the
emergence of closed time-like geodesics. Different models with superluminal propagation have been discussed, including nonlinear scalar field theories \cite{Blo82} noncommutative theories \cite{Noncomutative1}, waves in modifications of Einstein's theory with ether~\cite{Jacob}, and `superluminal' photons in the Drummond--Hathrell effect~\cite{Faster than gravity1,Ohkuma,plates1,UV2}.

The propagation of a object is called superluminal if it moves with a superluminal speed in the vacuum of ordinary quantum electrodynamics in an unbounded empty space. Arguments have been put forward that superluminal propagation in some cases can lead to causality paradoxes, for example, in the thought experiment with two black holes \cite{Dolgov} or with two plates in the Casimir effect, which move with high relative velocities. To avoid the emergence of closed time-like geodesics in such thought experiments, the authors of \cite{Liberati}, introduced the `chronology protection' \cite{Chronology} and showed that photons propagate in the effective metric that is different from the Minkowski metric. We note that superluminal propagation is not the only case where time-like geodesics can appear. There are several examples of GR space-times in which the local causality postulate is valid but closed time-like
geodesics nevertheless emerge \cite{Godel,Gott,Ori,otherCCC1,otherCCC2,otherCCC3,otherCCC3-4,
otherCCC4,otherCCC5,wormhole}.

We consider the $k$-essence defined by Lagrangian (\ref{sfaction}). Equations of motion (\ref{sfeom}) for the scalar field can be rewritten in the form
\begin{equation}
\tilde{\mathcal{G}}^{\mu \nu }\nabla _{\mu }\nabla _{\nu }\phi =0,
 \label{general eom}
\end{equation}
where the induced metric is defined as $\tilde{\mathcal{G}}^{\mu \nu }$ is defined as
\begin{equation}
\tilde{\mathcal{G}}^{\mu \nu } \equiv
\mathcal{L}_{,X}g^{\mu\nu}+\mathcal{L}_{,XX}\nabla^{\mu}\phi\nabla^{\nu}\phi,  \label{G2}
\end{equation}
This equation is hyperbolic, and its solutions are stable with respect to high-frequency perturbations if $(1+2X\mathcal{L}_{,XX}/\mathcal{L}_{,X})>0$ \cite{MukhGar,ArmenLim,Rendall}.
Small perturbations propagate in the effective metric $\mathcal{G}^{\mu\nu}$,
\begin{equation}
\mathcal{G}^{\mu\nu}\equiv\frac{c_{s}}{\mathcal{L}_{,X}^{2}} \tilde{\mathcal{G}}^{\mu \nu},
\end{equation}
where the sound speed $c_{s}$ is given by formula (\ref{sound in kessence}). Using the
matrix inverse to  $\mathcal{G}^{\mu\nu}$,
\begin{equation}
\mathcal{G}_{\mu\nu}^{-1}=\frac{\mathcal{L}_{,X}}{c_{s}}\left[g_{\mu\nu}
-c_{s}^{2}\left(\frac{\mathcal{L}_{,XX}}
{\mathcal{L}_{,X}}\right)\nabla_{\mu}\phi_{0}\nabla_{\nu}\phi_{0}\right],\label{Gdown}
\end{equation}
where $\left(...\right)_{,X}$ is a partial derivative with respect to $X$, we can find the induced metric interval
\begin{equation}
dS^{2}\equiv \mathcal{G}_{\mu\nu}^{-1}dx^{\mu}dx^{\nu},\label{interval}
\end{equation}
which determines the cone of influence of small perturbations of $k$-essence in the given background. (The indices are raised (lowered) using $g^{\mu\nu}$ ($g_{\mu\nu}$).) This influence cone is wider than the one determined by the metric $g_{\mu\nu}$, if
$\mathcal{L}_{,XX}/\mathcal{L}_{,X}<0$ \cite{ArmenLim,Rendall,Susskind,stringy causality1,stringy causality2,stringy causality3,stringy causality4}. As a result, superluminal propagation of small
perturbations is allowed. Here, we consider the $k$-essence as a secondary source of the gravitational field and ignore its back reaction on the metric.

\begin{figure}[t]
\begin{center}
\psfrag{XX}[t]{\LARGE$\mathbf{x'}$}
\psfrag{TT}[t]{\LARGE$\mathbf{t'}$}
\psfrag{X}[r]{\LARGE$\mathbf{x}$}
\psfrag{Y}[t]{\LARGE$\mathbf{y}$}
\psfrag{T}[t]{\LARGE$\mathbf{t}$}
\psfrag{XXX}[r]{\LARGE $\mathbf{\tilde x}$}

\includegraphics[angle=0,width=0.45\textwidth]{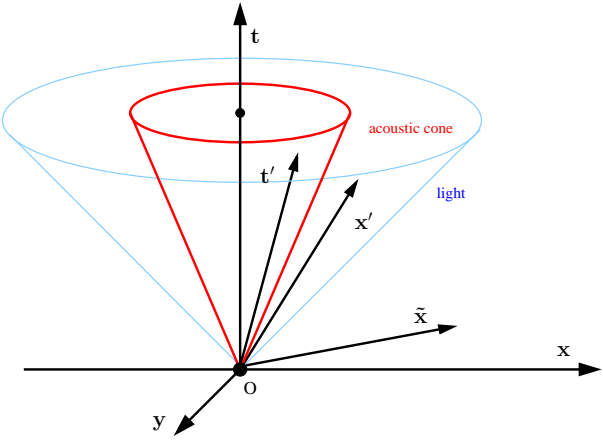}
\end{center}
\caption{How to create the `paradox' described above without using superluminal signals. The fluid is at rest, and sound perturbations are subluminal, $c_s<1$. The reference frame $(t',x')$ is obtained from the rest frame by a Lorentz transformation with the invariant velocity $c_s$. If the velocity of the moving reference frame exceeds the speed of sound, the hypersurface of constant time $t'$ lies inside the light cone, and the initial condition problem for the electromagnetic field is ill-posed. But if, instead of the frame  $(t',x')$, we use the `correct' frame  $(\tilde t,\tilde x)$, obtained from the rest frame by Lorentz transformations with the invariant speed of light, the
Cauchy problem is well-posed.}
 \label{soundlightcomes}
\end{figure}

\begin{figure}[t]
\begin{center}
\psfrag{T}[t]{\LARGE$\mathbf{t}$}
\psfrag{X}[t]{\LARGE$\mathbf{x}$}
\psfrag{XX}[l]{\LARGE$\mathbf{x'}$}
\psfrag{tt}[r]{\LARGE$\mathbf{t'}$}
\includegraphics[angle=0,width=0.45\textwidth]{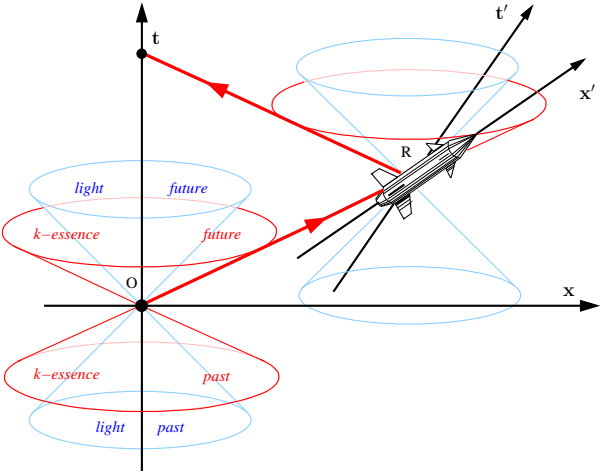}
\end{center}
\caption{The causal paradox does not arise when the superluminal signal propagates in a background violating the Lorentz symmetry (see Fig.~\ref{Bad Rocket}). The observer cannot send a signal to his own past.}
 \label{Good Rocket}
\end{figure}

\begin{figure}[t]
\begin{center}
\psfrag{T}[t]{\LARGE$\mathbf{t}$}
\psfrag{X}[t]{\LARGE$\mathbf{x}$}
\psfrag{XX}[l]{\LARGE$\mathbf{x'}$}
\psfrag{tt}[r]{\LARGE$\mathbf{t'}$}
\includegraphics[angle=0,width=0.45\textwidth]{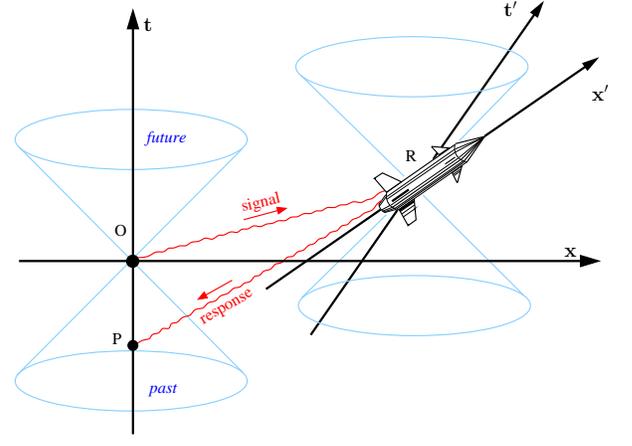}
\end{center}
\caption{Schematic representation of the tachyon causal paradox. Somebody living at the world line $x=0$, sends a tachyon signal in the direction $OR$ to an astronaut in a rapidly moving rocket $R$. In the rocket reference frame  $(x',t')$ the astronaut sends the tachyon signal back along the
trajectory $RP$. From the point of view of the astronaut, the signal $RP$ propagates in the direction of increasing $t'$, but moves back in time in the original reference frame. Hence, according to such a representation, it is possible to send signals to the past.}
 \label{Bad Rocket}
\end{figure}

We next discuss causality problems for the superluminal propagation of perturbations in the nontrivial background for different solutions, including the case of accretion of a noncanonical field onto a BH.

First of all, we consider the well-known paradox \cite{Tolman}, often referred to as the `tachyon anti-telephone', which arises when hypothetical superluminal particles --- tachyons propagating with a velocity $c_{\rm tachyon}>1$ --- are present. In this case, it is possible to send a signal to the past. Indeed, let some observer (see Fig.~\ref{soundlightcomes}), located at the point $x=0$ at rest in some reference frame $\left(x,t\right)$ send a tachyonic signal in the direction $OR$ to an astronaut moving in a rocket $R$. The astronaut, after having received the signal, sends the return tachyonic signal in the direction $RP$. The proper time of the astronaut $t^{\prime}$ increases while the signal is propagating. However, if the rocket speed exceeds $1/c_{\rm tachyon}$, the signal $RP$ propagates back in time in the original observer's reference frame. Thus, the observer in this case is able to

We now turn to the Minkowski space-time with a scalar field, which allows the superluminal propagation of perturbations in its background. For simplicity, we consider a time-dependent homogeneous field $\phi_{0}\left(t\right)$. The velocity $\partial_{\mu}\phi$ is directed along the space-like vector $u^{\mu}=(1,0,0,0)$. Why does a similar paradox not arise in this case? The point is that the superluminal propagation of signals (sound perturbations) is possible only in a nontrivial scalar field background. This background determines a preferential reference
frame, and the equation of motion for sound perturbations is no longer Lorentz invariant except in the special case $c_{s}=1$. In the moving astronaut's frame, equations for the perturbation propagation have a more complicated form than in the rest frame, and a special analysis is required to derive them. However, bearing in mind that signals in the k-essence propagate along the characteristics, which are coordinate-independent hypersurfaces, we can understand the character of propagation of signals in the astronaut's frame and in the rest frame to show that the signals always propagate forward in time in both frames (see Fig.~\ref{Good Rocket}). Hence, no closed time-like geodesics emerge in this case.

A note on the meaning of signals directed to the future and to the past should be made. As pointed out in \cite{Durrer2}, in order that no closed causal geodesics emerge in the $k$-essence during the superluminal propagation, the observer moving with a high speed relative to the background must send signals only in certain directions. However, we note that the notions of the future and the past are defined by the respective future and past cones, regardless of the particular choice of the reference frame. Thus, the signals forwarded to the future in the rest frame remain such in the reference frame of a fast moving rocket, in spite of the decrease in the time coordinate  $t^{\prime}$ for these signals. The contradiction appears because of a bad choice of the reference frame, due to which the decreasing time $t^{\prime}$ corresponds to signals forwarded to the future, and vice versa. An example that illustrates the above considerations is shown in Fig.~\ref{Bad Rocket}. Even without superluminal signals, the increase in the coordinate time does not imply the direction to the future.

Another question that frequently leads to misunderstanding is what speed must be considered as the propagation speed of signals: the phase velocity, the group velocity, or the velocity of the wave front. These points are discussed in \cite{BabMukVik08}, where, notably, it is shown that physical paradoxes do not arise during superluminal propagation of signals in a nontrivial background.

\subsection{Is it possible to look inside the black hole horizon?}
\label{lookinside}

It is of interest to study the behavior of noncanonical scalar fields near a BH, and we next consider the superluminal propagation in this case. First, by neglecting the back reaction of the fields on
the metric, we find the solution for stationary accretion onto a BH, and then we examine the propagation of perturbations in such a background.

We consider a scalar field with the Lagrangian density
\begin{equation}
\mathcal{L}(X)=\alpha ^{2}\left[ \sqrt{1+\frac{2X}{\alpha ^{2}}}-1\right] -\Lambda,
\label{Lagrange}
\end{equation}%
where $\alpha $ and $\Lambda $ are free parameters of the theory. The kinetic part of the action is the same as in \cite{MukhVik}, and for small values of derivatives in the limit $2X\ll \alpha ^{2}$the ordinary massless scalar field is recovered. It can be shown that no ghost solutions emerge in the theory with Lagrangian (\ref{Lagrange}).

As we already discussed in Section~\ref{teprichsec}, if the vector $\nabla_{\nu}\phi$ is
time-like (i.\,e., $X>0$ in our conventions), the field described by Lagrangian (\ref{Lagrange}), is formally equivalent to a perfect fluid with the density, pressure, and sound speed determined by
Eqs (\ref{e}), (\ref{p}) and (\ref{unublapsi}). Equation (\ref{sound in kessence}),
implies that the effective sound speed for a perturbation is
\begin{equation}
c_{s}^{2}\equiv \frac{\partial p}{\partial \varepsilon }=1+\frac{2X}{\alpha ^{2}}.
\label{soundspeed}
\end{equation}%
and for  $X>0$, it is always greater than the speed of light. In what follows, it is convenient to express the density and pressure in terms of this sound speed:
\begin{equation}
\varepsilon =\alpha ^{2}(1-c_{s}^{-1})+\Lambda ,~~p=\alpha ^{2}(c_{s}-1)-\Lambda .
\label{enpres_c}
\end{equation}%
It is easy to see that the null energy condition is satisfied, and therefore the Hawking theorem on the nondecreasing area of the BH horizon \cite{hawkell} is valid.

First of all, we find the stationary spherically symmetric background solution for the scalar field falling onto the BH. Here, we use the Eddington--Finkelstein coordinates with the metric
\begin{equation}
ds^{2}=f(r)dV^{2}-2dVdr-r^{2}d\Omega ,  \label{metric}
\end{equation}%
where $f(r)\equiv 1-r_{g}/r$, with $r_{g}\equiv 2M$ being the BH gravitational radius. The coordinate $V$ is related to the Schwarzschild coordinates $t$ and $r$ as $V\equiv t+r+r_{g}\ln |r/r_{g}-1|$. We assume that the accreting scalar field does not produce any back reaction on the metric. The stationarity condition suggests the following ansatz for the solution:
\begin{equation}
\phi (V,x)=\alpha \sqrt{c_{\infty }^{2}-1}\left( V+r_{g}\int F(x)dx\right) , \label{anz}
\end{equation}%
where $x\equiv r/M$ and $c_{\infty }$ is the sound speed at infinity. The common factor in
(\ref{anz}) is chosen so as to reproduce the cosmological solution at infinity, $\phi
(V,x)\rightarrow \alpha t\sqrt{c_{\infty }^{2}-1}$ and the factor $r_{g}$ in front of the integral is separated for convenience. The solution of (\ref{general eom}) that is not singular at the BH horizon is given by
\begin{equation}
F(x)=\frac{2}{f}\left( B\sqrt{\frac{c_{\infty }^{2}+f-1}{fx^{4}c_{\infty }^{8}/16+B^{2}\left(
c_{\infty }^{2}-1\right) }}-1\right) .  \label{F}
\end{equation}%
where $B$ is the integration constant to be determined below. The sound speed can then be found using Eqs (\ref{F}), (\ref{anz}) and (\ref{soundspeed}):
\begin{equation}
c_{s}^{2}=\frac{x^{3}c_{\infty }^{8}\left( xc_{\infty }^{2}/2-1\right) }{%
(x/2-1)x^{3}c_{\infty }^{8}+8B^{2}\left( c_{\infty }^{2}-1\right) }. \label{cs2}
\end{equation}%
We note that the sound speed becomes infinite at $x\equiv x_{sing}$, and this singularity is physical if there is a real solution of (\ref{F}) for all $x>x_{sing}$.

We now consider small perturbations on top of background (\ref{anz}), (\ref{F}). The characteristics (vector $\eta ^{\mu }$) for Eq (\ref{general eom}) satisfy the equations (see, e.\,g., \cite{ArmenLim,Rendall}):
\begin{equation}
\mathcal{G}_{\mu\nu}^{-1} \eta ^{\mu }\eta ^{\nu }=0,  \label{null}
\end{equation}
The vector $\eta^{\mu }$ describes the wave front propagation. It is possible to derive the following equation for the characteristics $\eta _{\pm }(x)\equiv dV/dx$:
\begin{equation}
\eta _{\pm }=\frac{1}{f}+\frac{1}{\xi _{\pm }},  \label{eta}
\end{equation}%
where
\begin{equation}
\xi _{\pm }=\pm f\sqrt{c_{\infty }^{2}-\frac{2}{x}}\,\frac{\sqrt{%
B^{2}(c_{\infty }^{2}-1)+c_{\infty }^{8}x^{4}f/16}}{c_{\infty }^{4}x^{2}f/4\mp B(c_{\infty
}^{2}-1)}.  \label{ksi}
\end{equation}%
We note that the equation $\xi _{\pm }=dx/dt$ describes the wave front propagation in the Schwarzschild coordinates $(t,x)$.

\begin{figure}[t]
\psfrag{V}[r]{\large$V\over r_{g}$} \psfrag{R}[t]{$r/r_{g}$} %
\psfrag{x}[t]{\small$x_{*}$} \includegraphics[width=0.45%
\textwidth,height=200pt]{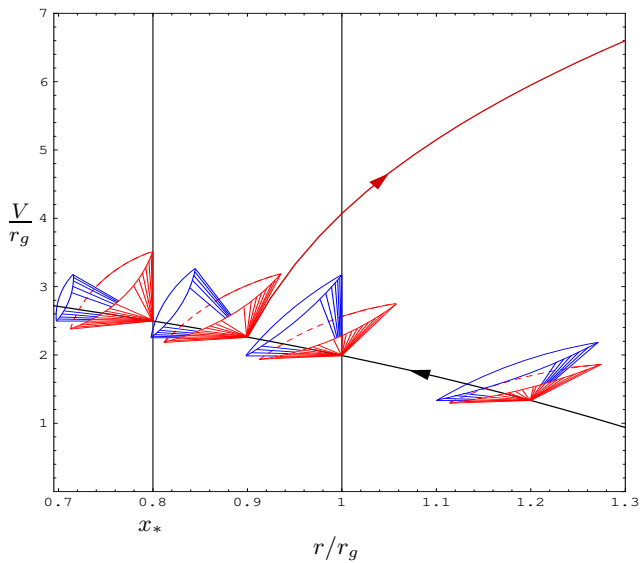}
\caption{Emission of a sound signal from a falling rocket in the Eddington--Finkelstein coordinates. The internal cones correspond to future light cones, and the external cones correspond to future sound cones (\protect\ref{eta}). The curve connecting the cones is a world line numerically
obtained from (\protect\ref{anz}) and (\protect\ref{F}) for the rocket comoving with the infalling photon field. When the rocket is located between the Schwarzschild ($r=r_{g}$) and the sound ($r=r_{\ast }$) horizons, a sound signal is emitted (along the curve marked with the arrow directed to the top right corner), which reaches the remote observer in a finite time interval. The signal trajectory is obtained by numerical integration of Eq (\protect\ref{eta}).} \label{cones}
\end{figure}

Equation (\ref{eta}) does not specify the propagation direction completely. In addition to the value of $dV/dx$, it is necessary to choose the future and past cones for each event. However, the location of the future and past light cones helps us to choose the sound cones. Using characteristics (\ref{eta}), we then determine the sound cones as follows: 1) the future and past sound cones do not intersect; 2) the future and past light cones are respectively contained inside the future and past sound cones. This is justified because this is the case at the spatial infinity, and sound characteristics (\ref{eta}) coincide there with radial light geodesics. As a result, we conclude that the signals propagating along $\eta _{+}$ and $\eta _{-}$, are respectively directed in the positive and negative  $V$-direction (see Fig.~\ref{cones}).

If the propagation vectors are known, we can find the location of the sound horizon. The sound horizon is defined as a surface on which the spatial velocity is equal to the sound speed. Signals that escape from the region above this surface can travel to the spatial infinity, and sound cannot come out from inside, because its propagation is limited by the superluminal motion of the fluid (as in the case of light capture by the event horizon in a gravitational field). The acoustic signal forwarded outside from the BH corresponds to $\eta _{+}$, and therefore the sound horizon is located at $x\equiv x_{\ast }$, where $\eta _{+}\equiv (dV/dx)_{+}$ becomes infinite (see Fig.~\ref{cones}).

We can now determine the integration constant $B$ in Eqs (\ref{anz}) and (\ref{F}). We assume that
in physically relevant cases, there must be no singularity at the sound horizon or outside it, as in the case of accretion of a perfect fluid. Hence, we have:

1) at $B\neq 1$1, either the physical singularity coincides with the sound horizon or the sound speed becomes imaginary (indicating absolute instability) inside some region outside the singular surface. In both cases, the solution is unphysical;

2) at $B=1$ and $c_{\infty }^{2}>4/3$, the sound speed becomes imaginary before reaching the sound horizon or singularity. This solution is also unphysical;

3) at $B=1$ and $c_{\infty }^{2}<4/3$, the sound horizon is located at $x_{\ast }=1/c_{\infty
}^{2}$, and the singularity is hidden inside the sound horizon. This is the only acceptable physical solution.

Therefore, we should set $B=1$ in (\ref{anz}), (\ref{F}), which completes the construction of the background solution. The accretion rate onto the BH is then expressed as
\begin{equation}
\dot{M}=4\pi M^{2}\alpha ^{2}(c_{\infty }^{2}-1)/c_{\infty }^{4}. \label{dotM}
\end{equation}

It is possible to show that for the found background solution, acoustic signals can indeed escape from the BH, which allows to look inside the BH using these signals. This becomes possible because, in the considered case, the sound horizon ($x_{\ast }=1/c_{\infty }^{2}$) is located inside the Schwarzschild horizon. As long as the signal is generated at sufficiently large $x$, namely, at $x>x_{\ast }$, it travels to the spatial infinity by propagating along $\eta _{+}$. For example, at the event horizon,
\begin{equation}
\eta _{\pm H}=\frac{1}{2}\frac{\left( c_{\infty }^{4}\pm 1\right) ^{2}}{%
c_{\infty }^{2}-1}.
\end{equation}%
The propagation vector $\eta _{+H}$ is positive, and therefore the signal can freely escape from the Schwarzschild horizon and escape the BH. Fig.~\ref{cones} illustrates the way the signal escapes from a BH.

The main result in this section is that in the case of accretion of a special Born--Infeld-like field onto a BH, the information can be sent from inside the BH horizon to the outside. This result has a purely classical nature. It also changes the usual concept of the event horizon of a BH as the absolute barrier for outward motion. Here, the cosmic censorship principle (see Section~\ref{extreme}) is not violated, because the central singularity is hidden under the sound horizon. The  null energy condition in this model is also satisfied, and the BH mass during accretion does not decrease. A similar (in some sense) possibility of escaping from a BH also occurs in bimetric theories \cite{Bimetric}.

\subsection{Ghost condensate in the black hole field}
\label{condensate}

The ghost condensate is a scalar field theory with the k-essence Lagrangian and additional higher-order derivative terms. (See \cite{ghost} for a discussion of the ghost condensate.) The $k$-essence part can be taken in the form
\begin{equation}
\label{GC}
\mathcal{L} =\frac{1}{2\tilde M^4}\left(X-\tilde M^4\right)^2.
\end{equation}
It can be shown that for small $X$, namely, $X<\tilde M^4$, this theory contains a ghost, while for $X>\tilde M^4$ ghosts do not appear. This is why this theory is referred to as `ghost condensate'. It is assumed that the ghost condensate is an effective field theory considered at $X\sim \tilde M^4$. The cosmological evolution implies that $X\to \tilde M^4$  for a homogeneous solution. It is easy to see from (\ref{p}) and (\ref{e}) that this solution corresponds to $\rho(X)\to 0$ and $p(X)\to 0$. Thus, a nontrivial field configuration $\dot\phi_c = const$, appears, but the energy and pressure in this homogeneous solution vanish.

Formulas (\ref{cpoint1}) allow calculating that the critical point is located at the radius $3r_g$, and here $\tilde A = 1$ and, accordingly, the BH mass increases at the rate $\dot M = 2\pi \tilde M^4 r_g^2$. Such a calculation was first carried out in \cite{Fro04}. A paradoxical situation arises: the ghost condensate field does not contribute to the cosmological evolution (because, as noted above, the pressure and density at the point of interest, $X\to \tilde M^4$ are zero), while the accretion rate is nonzero; moreover, it is proportional to $\tilde M^4$. Therefore, by a special choice of $\tilde M$ (we recall that $\tilde M$ is the parameter in the Lagrangian), the energy flux can be made very large, while leaving the cosmological density and pressure equal to zero.

Using Eq (\ref{evol1}) it is possible to explain why this situation arises for the ghost condensate. The dimensionless accretion parameter $\tilde A$ is usually of the order of unity, and $\dot\phi_\infty$ is then unrelated to pressure and density, while, for example, in the case of the canonical field, $\dot\phi_\infty$ uniquely determines density and pressure. In fact, such a strange behavior of accretion is due to the pathological behavior of scalar field (\ref{GC}) at the point $X=\tilde M^4$: the ghost condensate behaves like dust at this point. That is why additional terms with higher-order derivatives were originally included in the ghost condensate Lagrangian. This pathology can also be seen from another accretion equation, Eq (\ref{evol}), which implies that $\dot M =0$, if $\rho_\infty = p_\infty =0$. This result contradicts what we have just obtained from the critical point consideration, but in fact it corresponds to the choice of another branch of the solution. Due to the dust-like behavior of the ghost condensate, we can choose branches of the solution that have no critical point (supersonic branches). In particular, it is possible to explicitly write the solution for $\phi$ (see \cite{Fro04}) with the vanishing actual flux, similarly to what is obtained from (\ref{evol}). Therefore, this paradox is resolved by the `correct' choice of the physical solution, namely, the one that arises as a result of the evolution. A more detailed analysis of the ghost condensate accretion onto a Schwarzschild BH taking higher-order derivative terms at the point $X\simeq \tilde M^4$, into account (which are necessary to make the theory regular at this point) can be found in \cite{Mukohu}. The accretion rate was found to be very small.

We consider another paradox related to the motion of the ghost condensate in the gravitational field of a BH. As shown in \cite{Dubovsky:2006vk} it is possible to violate the second law of thermodynamics using the ghost condensate. Namely, it is possible to obtain the event `horizons' with different temperatures for one state of a BH (in particular, with a zero accretion rate). To
create the second event horizon around a BH, in addition to the ghost condensate field, it is necessary to introduce another field kinetically connected with the first field, such that the
effective metric of the second field is different from the standard one (in analogy with the situation in the case of $k$-essence).

We consider the ghost condensate described by (\ref{GC}), and in addition introduce the field described by the Lagrangian \cite{Dubovsky:2006vk}
\begin{equation}
\L_\psi = \frac12(\partial_\mu\psi)^2 + \frac{\varepsilon}{2}(\partial_\mu\phi \partial^\mu\psi)^2,
\end{equation}
where $\varepsilon$ is a parameter. It is then possible to find a regular solution for the ghost condensate, such that $X = M^4$ (for this, it is necessary to pass to the coordinate frame regular on the horizon and to assume that $\phi$ is proportional to the time in these coordinates). Because the solution for $\phi$ is nontrivial, the metric for the $\phi$ field is nonstandard, with the signal (small perturbation) propagation velocity $v= 1/(1+\epsilon)^{1/2}$. By choosing positive or negative values of $\epsilon$, it is possible to obtain subluminal or superluminal propagation of signals for $\psi$. Hence, we can find that the Hawking radiation for particles is $\psi = v^3 T_H$, where $T_H$ is the Hawking radiation for ordinary particles (for example, gravitons). The mere fact that another (induced) metric and another temperature exist is not astonishing: we have seen that a similar situation occurs for the $k$-essence. Moreover, phonons in an ordinary fluid also propagate in the induced metric background (with the exception of an ultrahard fluid). Another aspect is interesting here: no accretion of matter occurs in this case (because the energy flux for the considered solution vanishes), and the state of the BH does not change, i.\,e., the BH mass does not increase! In other words, we have a situation in which a body (the BH in this case) emits different particles (for example, gravitons and the field $\psi$) with different temperatures, but the state of the body does not change if the Hawking radiation can be neglected.

Such a system violates the second law of thermodynamics \cite{Dubovsky:2006vk}. Indeed, let the Hawking temperatures of two particles be $T_2$ and $T_1$, and $T_2 < T_1$ for definiteness. We encircle the BH by two shells: one (shell $A$) interacts only with particles $\psi_1$, and the other (shell $B$) interacts only with particles $\psi_2$. We take the temperatures $T_A$ and $T_B$ of these shells to be such that $T_1<T_A < T_B < T_2$. Clearly, the net flux of particles $1$ is directed toward the BH, while the net flux of particles $2$ is directed outward from the BH. It is possible to find temperatures $T_A$ and $T_B$ such that these fluxes are equal (not violating the previous condition). Then the total flux onto the BH vanishes, but the heat flows (through the BH) from the colder shell (with the temperature $T_A$) to the hotter shell (with the temperature $T_B$).

\subsection{Galileon accretion}

Scalar and scalar-tensor models, which are currently widely known under the name of Galileon, have been studied in physics and mathematics in different `reincarnations'. The nonlinear fourth-order partial differential equation (now called the Monge--Ampere equation) applied to different problems of Riemann geometry, conformal geometry, and so on was investigated as early as the 18th century. In 1974, Horndeski formulated the most general scalar-tensor theory in four dimensions, whose equations of motion include derivatives of the order not higher than two \cite{Horndeski}. Then, in the 1990s, Fairlie et al. \cite{Fairlie:1991qe} developed the so-called universal field theory, which is constructed step-by-step: the next Lagrangian is determined from the equations of motion of
the previous one. Recently, the model known as the Galileon \cite{Nicolis:2008in}, which has been further elaborated in many papers (see, e.\,g., \cite{Deffayet:2009wt,Deffayet:2009mn,Deffayet:2010zh}. A remarkable property of this theory is that its Lagrangian has higher-order terms, but only derivatives of the second order and below enter the equations of motion.

The Galileon model is interesting in several respects. First, this is a theory with a nonquadratic kinetic coupling, which leads to the propagation of perturbations in the effective metric different from the gravitational one, as in the case of $k$-essence. Another interesting feature of the Galileon is the possibility to reproduce the cosmological model with phantom behavior, but without ghost solutions for some parameters and initial conditions.

We consider the spherically symmetric accretion of a Galileon onto a Schwarzschild BH in the test fluid approximation. We assume that the Galileon evolves on a cosmological time-scale. The general form of the covariant action for the Galileon as a scalar field is given by
\cite{Deffayet:2009wt}
\begin{equation}
\label{action0} S_\pi = \int d^4 x\sqrt{-g}\,\L_\pi,
\end{equation}
where the Lagrangian density can be represented as the linear combination
\begin{equation}
\label{L} \L_\pi = \sum_{i=1}^{i=5} c_i\L_i,
\end{equation}
with
\begin{equation}
\label{L1234}
\begin{aligned}
\L_1 &= \pi, \;\; \L_2 = \pi_{;\mu} \pi^{;\mu}, \;\;
\L_3 =  \pi_{;\mu} \pi^{;\mu} \Box\pi, \\
\end{aligned}
\end{equation}
The terms $\L_4$ and $\L_5$, which have a more complicated structure and contain higher-order derivatives of $\pi$ are not considered here. We also set $c_1=0$, i.\,e., we exclude the
`potential' term.

To simplify formulas, it is convenient to introduce dimensionless variables
\begin{equation}
\label{rescale} x^\mu\to r_g x^\mu,\quad  \quad \pi \to Cr_g \pi,
\end{equation}
where $r_g=2M$ is the BH gravitational radius and the constant $C$ can be associated with the cosmological quantity $\partial_t\pi$.

We examine accretion of a Galileon with nonzero $\L_2$ and $\L_3$, and other terms set to zero. This type of action (up to coefficients in front of $\L_2$ è $\L_3$) appears in the effective actions for the scalar field in a certain limit of the Dvali--Gabadadze--Porrati (DGP) model \cite{Dvali:2000hr}:
\be
 \label{L2L3action}
  S_\pi = r_g^4 C^2\int d^4
x\sqrt{-g}\,\left[ \e \left(\partial\pi\right)^2 + \kappa \left(\partial\pi\right)^2\Box\pi
\right],
\ee
where $\e = 0,\, \pm1$, $\kappa = C c_3/r_g$, and we allow both positive and negative values of  $\kappa$. Positive $\e$ correspond to the canonical kinetic term, and positive $\e$ and $\kappa$ yield the Lagrangian of the DGP scalar field.

The equations of motion obtained from (\ref{L2L3action}), are
\begin{equation}
\label{eom}
 \nabla_\mu j^\mu = 0,\quad j_{\mu} \equiv 2 \e\, \pi_{,\m} + \kappa\, \left(2
\pi_{,\m}\Box\pi - \partial_\mu \left(\partial\pi\right)^2\right),
\ee
or
\begin{equation}
\label{eom1} \e\,\Box\pi +\kappa\left( \left(\Box\pi\right)^2 - \left(\nabla\nabla\pi\right)^2
-  R^{\m\n}\pi{,_\m}\pi_{,\n} \right)= 0.
\end{equation}
We also need the equation for perturbations $\delta\pi$ in the nontrivial background $\pi(t,x)$ in the high-frequency limit. From (\ref{eom1}), we obtain the equation
\be
 \label{eomG}
G^{\mu\nu}\nabla_\mu\nabla_\nu\d\pi= 0,
\ee
where
\be
 \label{G} G^{\m\n} = \left(\e+2\kappa\Box\pi\right)g^{\m\n}
 - 2\kappa \nabla^\m\nabla^\n\pi.
\end{equation}
The propagation vector for small perturbations can be found from the relation
\begin{equation}
\label{vectors} {G}^{-1}_{\mu\nu}\eta^\mu\eta^\nu = 0,
\ee
where ${G}^{-1}_{\mu\nu}$ is the matrix inverse to $G^{\mu\nu}$.

Because we are interested in solutions for the scalar field, in some region inside the Schwarzschild horizon in particular, we use the Eddington--Finkelstein coordinates, which are regular at the horizon. The Eddington--Finkelstein coordinates $(v,\, r)$ are connected with the ordinary Schwarzschild coordinates $(t,\, r)$ by the relation
\begin{equation}
\label{EFcoord} v= t+ \int\frac{dr}{f},\quad r=r,\nonumber
\end{equation}
where $f=1-1/r$ in dimensionless variables. The Schwarzschild metric in the Eddington--Finkelstein coordinates becomes
\begin{equation}
\label{EF} d s^2 =  f d v^2 - 2d vdr - r^2 d \Omega,
\end{equation}
To study stationary accretion, we use the ansatz
\begin{equation}
\label{ansatz} \pi(v,r) = v - \int\frac{dr}{f} + \psi(r).
\end{equation}
We note that in adopting ansatz (\ref{ansatz}), we have freedom in choosing the normalization in
(\ref{rescale}). Thus, we can take the constant $C$ equal to $\partial_t\pi$ at the spatial infinity,
\begin{equation}
\label{C} C = \partial_t\pi |_{r=\infty} = \partial_v\pi|_{r=\infty},\nonumber
\end{equation}
thereby setting the coefficient $v$ in (\ref{ansatz}) equals to unity. Because the current depends only on $r$, Eq (\ref{eom}) can be integrated once. As a result, we obtain
\begin{equation}
\label{1int} r^2 j^r = A,
\end{equation}
where $A$ is a constant that determines the total flux. For ansatz (\ref{ansatz}) the  $r$-component of the current takes the form
\begin{equation}
\label{jr} j^r =  2\e\, f \psi '+ \kappa  \left(-\frac{f'}{f}+f f' \psi '^2+\frac{4 f^2 \psi
'^2}{r}\right).
\end{equation}
Equations (\ref{1int}) and (\ref{jr}) can also be derived from $T^{\mu\nu}_{\hphantom{\mu;};\nu}=0$, which yields $r^2T_v^r = {\rm const}$, whence for ansatz (\ref{ansatz}) we find $T_v^r = j^r$. Equations (\ref{1int}) and (\ref{jr}) yield an algebraic equation for $\psi'$. The solution contains a free parameter $A$: $ \psi' = \psi'(A,r)$. The physical solution is obtained from the condition of the absence of singularities at the Schwarzschild horizon and the sound horizon. In general, Eqs (\ref{1int}) and (\ref{jr}) have two solutions:
\begin{equation}
\label{sol12} \psi'_{(2,3)}=
 - \frac{ \e\,r^2 f\pm
 \sqrt{\e^2\, r^4 f^2\! +\!\kappa r \left(A f\!
  +\!\kappa r^2 f' \right)\left(r f'\!+\!4 f\right) }}
 { \kappa r f \left(r f'+4 f\right)},
\end{equation}
where the index $(2,3)$ means that the solution is obtained in the theory with the $\L_2$ and $\L_3$ terms in the Lagrangian. Solutions $\pi'(r)$ for different values $\epsilon$ and $\kappa$ are shown in Fig.~\ref{fig DGP} (see also \cite{Bab11}).

\begin{figure}[t]
\begin{center}
\includegraphics[width=0.45\textwidth]{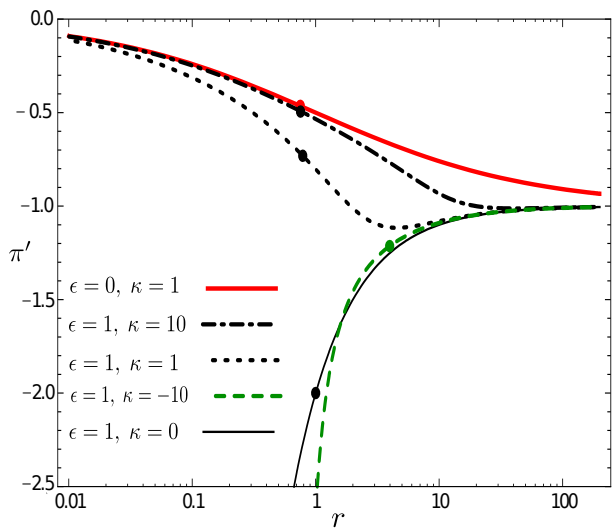}
\caption{Solutions $\pi' = -1/f+\psi'(r)$ for different model parameters in the case
$\L\sim \L_2 +  \L_3$. The locations of the sound horizon are shown by dots. At $\kappa <0$ and $\kappa>0$, the sound horizon respectively lies outside and inside the Schwarzschild horizon.} \label{fig DGP}
\end{center}
\end{figure}

Because we are considering the problem in the stationary case for a test fluid, the rate of the BH mass change can be found from the total flux at infinity, $r\to \infty$. In the Schwarzschild coordinates, the total flux $\propto (r^2 T_t^r)$. Expressing $T_t^r$ through components in the Eddington--Finkelstein coordinates, we obtain
\begin{equation}
\label{BHmass}
\frac{dM}{dt} =  4\pi A\, r_g^2 \dot{\pi}^2_\infty.
\end{equation}
In the final expression (\ref{BHmass}), we changed back to physical units. The flux can be made negative by changing the common sign of the Lagrangian for $\pi$, and the BH mass then decreases.
Usually, this sign change is associated with the appearance of ghost solutions. However, it was shown in \cite{Deffayet:2010qz} that although the term $\L_2$ has a `ghost'-like form, the total Lagrangian $\L\sim\L_2 + \L_3$ does not have ghosts near the cosmological attractor.

\section{Accretion with back reaction}
\label{backreaction}

In the models considered in Sections~\ref{perfect}--\ref{condensate}, accretion was
considered in the test fluid approximation. This means that the fluid `felt' the gravitational field of a BH and moved in a given external gravitational field, and the gravitational field of the fluid itself was ignored. But the gravitational field of the fluid sometimes becomes fundamentally important and can qualitatively change the process. The field of the accreting fluid and related phenomena are referred to as back reaction effects. In this section, we study the back reaction of the accreting matter on a spherically symmetric BH using methods of the theory of perturbations in the stationary accretion case.

\subsection{Approaching the extreme state and shortcoming of the test fluid model}
\label{extreme}

In the accretion of a phantom fluid with $\rho+p<0$, the Reissner--Nordstr\"om BH mass decreases. The question arises as to whether this process allows transforming a Reissner--Nordstr\"om BH into a naked singularity. If the back reaction effects are neglected, such a transformation seems plausible, because the BH mass decreases, while the electric charge is conserved. The transformation of a
Reissner--Nordstr\"om BH into a naked singularity through the accretion of a fluid with $\rho<0$ was discussed in \cite{DorHan10,Sch10}.

The transformation of a BH into a naked singularity means the violation of the cosmic censorship principle. This principle was formulated by Penrose \cite{penrose69} in 1969 on the basis of theorems about singularities in GR \cite{penrose65,penrose68,hawkell} and the general properties of BHs. The cosmic censorship principle states that for any physical process, the central singularity remains hidden from a remote observer by the BH event horizon. Notably, a BH --- and not a naked singularity --- is always formed during gravitational collapse. The cosmic censorship principle remains unproved and is only a plausible hypothesis \cite{penrose73,Pen73,wald74,israel84}. This principle underlies the third law of BH thermodynamics \cite{bch73}, which states that it is impossible to reach the extreme state of a BH and, accordingly, to transform the BH into a naked singularity in a finite number of steps. The cosmic censorship principle was verified for electrically charged and rotating BHs in the test particle approximation \cite{bardeen69,wald97,barausse10,japan11,bardeen70,roman88}. Well-known examples of the cosmic censorship principle violation have been realized under extremely unphysical conditions of matter collapse with an unrealistic strongly anisotropic energy-momentum tensor.
The decrease in the black hole mass via phantom energy accretion opens up the principal possibility of violating the third law of BH thermodynamics in the case where the BH rotates or has an electrical charge. The charge and angular momentum conservation during such accretion allow an extreme state to be reached in a finite period of time. According to this logic, if accretion continues, the event horizon should disappear, and the BH must transform into a naked singularity. We note that this possibility is realized in the test fluid approximation. In Sections~\ref{corrections} and \ref{shellssec}, we argue (but do not prove) that the third law of BH thermodynamics remains valid during the phantom energy accretion if the back reaction of the accreting matter on the metric of an almost extreme BH is taken into account.

We always assumed in the foregoing that the fluid has no back reaction on the metric. This approximation fails for nearly extreme BHs. The presence of an arbitrary light fluid can dramatically change the metric, and the back reaction of the accreting matter can prevent the BH from converting into a naked singularity. The possibility of back reaction in such problems was considered in \cite{hod08} in the context of the absorption of scalar particles with large angular momenta by a nearly extreme BH.

In \cite{Babetal08}, accretion onto an extreme BH was studied. At the BH event horizon with $r_+=M$, the radial component of the 4-velocity, $u^r$, was shown to tend to zero, $u^r\to 0$ and the density $\rho$ to infinity, $\rho\propto n^2\propto(r-M)^{-1}\to\infty$. The total mass of the fluid near the BH also increases infinitely. Such a behavior signals a violation of the test fluid approximation. For this reason, the obtained solution is not fully self-consistent, and to obtain correct solutions, the back reaction effects should be considered.

\subsection{Perturbation theory and corrections to the metric}
\label{corrections}

In this section, we present the perturbative method for the accretion problem \cite{BabDokEro12}, which in the first approximation takes the back reaction effects into account. We find corrections to the metric that appear due to the influence of accreting matter with an energy-momentum tensor of the general form $T_{\mu\nu}$ on the metric by assuming quasi-stationary accretion, when the accretion rate onto a BH is small. The appearance of a small parameter in the case of quasi-stationary accretion significantly facilitates the calculation of the back reaction effects on the metric. In this case, in the Einstein equations, we can neglect both second-order derivative terms and products of the first-order derivative terms, which are quadratic in the small parameter. The solution for  $T_{\mu\nu}(r)$ on the metric by assuming quasi-stationary accretion, when the accretion rate onto a BH is small. The appearance of a small parameter in the case of quasi-stationary accretion significantly facilitates the calculation of the back reaction effects on the metric. In this case, in the Einstein equations, we can neglect both second-order derivative terms and products of the first-order derivative terms, which are quadratic in the small parameter. The solution for $T_{\mu\nu}(r)$ as a function of radius is assumed to be known from the test-fluid approximation, i.\,e., when ignoring the back reaction effects. Then, using $T_{\mu\nu}(r)$ as the zeroth approximation, we find corrections to the metric due to the back reaction.

We first apply this scheme to a Schwarzschild BH in the free-falling reference frame. A spherically symmetric metric can be written in the form (see, e.\,g., \cite{Bondi64,LL}):
\begin{equation}
    \label{EF2}
    ds^2 = e^{\nu(V,r)+2\lambda(V,r)}dV^2 - 2 e^{\lambda(V,r)}dVdr -r^2d\Omega,
\end{equation}
where $\nu(V,r)$ and $\lambda(V,r)$ are arbitrary functions. This reference frame is by construction similar to the Eddington--Finkelstein frame and is connected with radially falling photons (null geodesics). A similar metric was introduced in \cite{radiationcoordinates} using `radial coordinates'. The vacuum Schwarzschild solution is recovered by setting $\lambda =0$ and $e^{\nu(V,r)} = 1- 2M_0/r $, where $M_0$ is the BH mass, whence
\begin{equation}
    \label{EF0}
    ds^2_\text{vac} = \left( 1 - \frac{2M_0}{r} \right) \, dV^2 - 2 dVdr -r^2d\Omega,
\end{equation}
We note that Eq (\ref{EF0}) is a solution for the metric in the zeroth approximation.

Similarly, instead of the metric coefficient $\nu(V,r)$, we use the function $M(V,r)$,
defined as
\begin{equation}
    \label{corr}
    e^{\nu(V,r)}  \equiv 1-\frac{2M(V,r)}{r},
\end{equation}
such that $M(V,r) = M_0={\rm const}$ in the zeroth approximation.

Substituting (\ref{corr}) and (\ref{EF2}) in the Einstein equations, we arrive at the system of equations:
\begin{eqnarray}
        8\pi T_0^{\phantom{0}0} &=&  -e^\nu\left(\frac{1}{r^2}+\frac{\nu'}{r}\right)+\frac{1}{r^2}  , \label{EG00}\\
     8\pi T_0^{\phantom{0}1}  &=&   \frac{e^\nu}{r}\dot\nu,       \label{EG01}    \\
    8\pi T_1^{\phantom{0}0} &=&   \frac{2\left(e^{-\lambda}\right)'}{r}, \label{EG10}\\
    8\pi T_1^{\phantom{0}1}   &=& -e^\nu\left(\frac{1}{r^2}+\frac{\nu'}{r}\right)+\frac{1}{r^2} - \frac{2\lambda'}{r}e^\nu,  \label{EG11}\\
    8\pi T_2^{\phantom{0}2}     &=& 8\pi T_3^{\phantom{0}3} =  -e^\nu\left(\lambda''+\frac{\nu''}{2}\right)    -e^{-\lambda}\dot{\lambda}' \nonumber \\
    & &- e^\nu\left(\lambda'^2+\frac{\nu'^2}{2}+\frac{\lambda'+\nu'}{r} +\frac32\lambda'\nu'\right),  \label{EG22}
\end{eqnarray}
where the dot denotes $\partial/\partial V$, and the prime denotes $\partial/\partial r$. The
left-hand sides of Eqs (\ref{EG00}-\ref{EG22}) contain the energy-momentum tensor components taken in the zeroth approximation, i.\,e., the solution for the stationary accretion of matter in which back reaction is ignored. Not all equations in the system (\ref{EG00})--(\ref{EG22}) are independent. Using the Bianchi identity, it can be shown that (\ref{EG22}) is a combination of Eqs (\ref{EG00})--(\ref{EG11}).

Substituting (\ref{corr}) in (\ref{EG00}) and (\ref{EG01}), we obtain
\begin{equation}
\label{Mprim} M' = 4\pi T_0^{\phantom{0}0} r^2,
\end{equation}
\begin{equation}
\label{Mdot} \dot M =  {\cal A}.
\end{equation}
In Eq (\ref{Mdot}), we introduce the notation
\be
 \label{flux}
 {\cal A} \equiv
- 4\pi T_0^{\phantom{0}1} r^2,
\ee
for the total energy flux crossing the surface of a radius $r$. The right-hand sides of Eqs (\ref{Mprim}) and (\ref{Mdot}) are taken in the zeroth approximation. The energy-momentum tensor components in this approximation are independent of time, and the flux, ${\cal A}$, is independent of $r$, i.\,e., ${\cal A} =$const.

Integrating (\ref{Mprim}) and (\ref{Mdot}), we find
\bea \label{master1}
    M(V,r) &=& M_0 + \cA V+ 4\pi \int_{r_0}^r T_0^{\phantom{0}0}(r)r^2dr, \\
\label{master2}
    \lambda(r) &=& - 4\pi \int_{r_0}^r T_1^{\phantom{0}0} rdr.
\eea
If the energy-momentum tensor components are sufficiently smoothly varying functions of the radial coordinate (which seems to be quite reasonable for nonpathological matter), then from
(\ref{master1}) and (\ref{master2}) it is possible to find corrections to the metric near the BH horizon:
\bea \label{m1hor}
    M(V,r) &=& M_0\! +\! \cA V\!+\! 4\pi r_0^2\,  \left(r\!-\! r_0\right) T_0^{\phantom{0}0}\mid_{r=r_0}, \\
\label{m2hor}
    \lambda(r) &=& - 4\pi r_0 \left(r- r_0\right) T_1^{\phantom{0}0}\mid_{r=r_0}.
\eea
The obtained results can easily be generalized to a Reissner--Nordstrom BH, if instead of (\ref{corr}) we write the metric coefficient $\nu$ in the form
\begin{equation}
    \label{nuRN}
    e^{\nu(V,r)} = 1-\frac{2M(V,r)}{r} + \frac{Q^2}{r^2},
\end{equation}
where $Q$ is the BH charge.

We calculate the shift of the apparent horizon (or visibility horizon) due to the corrections found above. For metric (\ref{EF2}) the location of the apparent horizon $r_h$ can be found from the equation (see e.\,g., \cite{Nielsen:2010gm})
\begin{equation}
    \label{corr0}
    e^{\nu(V,r)} = 0.
\end{equation}
This equation can be obtained from the condition that $dr/dV =0$ for radially moving photons. Indeed, from $ds^2=0$, we find two radial null geodesics:
\begin{equation}
dV=0, \quad \frac{dr}{dV}e^{-\nu-\lambda}=\frac{1}{2}. \label{lgeo2}
\end{equation}
The apparent horizon satisfies the condition that photons do not cross the surface $r=const$ for increasing $r$. This yields $dr/dV=0$, and from (\ref{lgeo2}), we obtain $e^{\nu+\lambda}=0$; hence, $e^{\nu(V,r)}=0$. The last transition is based on the regularity of the function $\lambda$. For a Schwarzschild BH, from (\ref{corr0}) and (\ref{master1}), we now derive an explicit equation for $r_h$:
\begin{equation}
\label{corr2}
 M_0 + \cA V+ 4\pi \int_{r_0}^{r_h} T_0^{\phantom{0}0}(r)r^2dr  = \frac{r_h}{2}.
\end{equation}
For small shifts of the horizon,
\begin{equation}
r_h \approx 2M_0 + 2 {\cal A} V.
\end{equation}
and does not depend on other components of the energy-momentum tensor.

Similarly, for the horizon shift of a charged BH, instead of (\ref{corr2}) we find
\begin{equation}
\label{corr3}
 M_0 + \cA V+ 4\pi \int_{r_0}^{r_h} T_0^{\phantom{0}0}(r)r^2dr
 = \frac{r_h}{2} + \frac{Q^2}{2 r_h},
\end{equation}
\begin{equation}
\label{rhRN} r_h \approx
M_0 + {\cal A} V +\sqrt{M_0^2 - Q^2 + 2 M_0 \cA V}.
\end{equation}
In the case of accretion of phantom energy, the horizon does not exist for positive $V$, which means that our method of the theory of small perturbations cannot be applied here. This is because the phantom energy accretion decreases the BH mass and an arbitrarily small amount of phantom energy can transform the BH into a naked singularity, and such a transformation cannot be described in terms of the quasi-stationary approximation that we use. On the other hand, if normal (not phantom) matter is accreted, Eq (\ref{rhRN}) is fully applicable. We note that in static coordinates, the test-fluid
approximation is violated during accretion of any type of matter (either phantom or non-phantom) \cite{DokEro11}.

It is also interesting to note that near the horizon of a Schwarzschild BH, irrespective of the form of the energy-momentum tensor of the accreting fluid, the metric has the Vaidya solution form \cite{Vaidya1}.

We now consider the accretion of a perfect fluid with energy-momentum tensor (\ref{emtensor}). In the reference frame used here, the 4-velocity has the form
\begin{equation}
    \label{u}
    u^\mu = \left(\frac{1}{\sqrt{f_0+u^2}+u},\; -u,\; 0,\; 0\right),
\end{equation}
where $u\equiv |dr/ds| >0$ is the absolute value of the radial component of the 4-velocity in the static coordinates and $f_0\equiv 1-2 M_0/r$. It easy to verify that the components $u^\mu$ and $u_\mu$ do not diverge at the horizon. The corresponding components of the energy-momentum tensor are expressed as
\be
T_0^{\phantom{0}0} =  \frac{\rho\sqrt{f_0+u^2}-pu}{\sqrt{f_0+u^2}+u} ,\label{EMT00} \quad
T_1^{\phantom{0}0} = -\frac{\rho+p}{\left(\sqrt{f_0 + u^2}+u\right)^2}. \ee
Near the horizon, $f_0\to 0$, and we obtain
\be
    T_0^{\phantom{0}0}  \to \frac{1}{2}(\rho-p),\quad
    T_1^{\phantom{0}0} \to -\frac{\rho+p}{4u^2}.
\ee
Thus, from (\ref{m1hor}) and (\ref{m2hor}), it is possible to calculate corrections to the metric in the form
\bea
M(V,r) &\approx& M_0+ \cA V+ 2\pi r_0^2 \left(\rho - p\right)\left( r-r_0 \right),
 \label{midealfl} \\
\lambda (r)&\approx&   \pi r_0 \frac{\rho+p}{u^2}\left( r-r_0 \right).
\label{lambda1}
\eea
Expressions (\ref{midealfl}) and (\ref{lambda1}) are valid for any perfect fluid near the BH horizon. We note that the energy flux onto the BH is given by expression (\ref{flux}) with
$T_0^{\phantom{0}1}=-(\rho+p)u\sqrt{f_0+u^2}$ for a perfect fluid \cite{BabDokEro04}. Therefore, it is evident from (\ref{midealfl}) that the accretion of phantom energy with $\rho+p<0$ leads to a
decrease in the BH mass. Therefore, we have confirmed that taking back reaction into account does not change the result obtained in Section~\ref{idealsec}, where only the zeroth approximation
was analyzed.

A similar calculation carried out in the static reference frame \cite{DokEro11} showed that the test fluid approximation is violated due to the back reaction of the fluid gravity on the metric when accretion occurs onto a BH approaching the extreme state $M\to Q$. Namely, corrections to the BH event horizon and the internal Cauchy horizon diverge for arbitrarily small accretion rates $\dot M$ as $M\to Q$. This conclusion is in agreement with the cosmic censorship principle \cite{penrose69} and the third law of BH thermodynamics \cite{bardeen73}, according to which the extreme state is unreachable by finite processes. In other words, it is impossible to transform a BH into a naked
singularity by such processes. However, to fully clarify the back reaction mechanisms, an analysis of the full nonlinear Einstein equations is required.

\subsection{Accretion of thin shells}
\label{shellssec}

Models of a thin self-gravitating shell can also be used to investigate the back reaction of accreting matter on the metric. Although a medium with nonzero pressure cannot be fully represented in the thin shell model, this approach turns out to be very useful because it allows making precise conclusions about the global structure of a spherically symmetric space-time. Models with shells can be quantized, and the corresponding solutions prove to be useful in understanding the Hawking radiation mechanism. The shell models have been applied not only to solving problems with BHs but also to calculating phase transitions in the early Universe.

An elegant generally covariant formalism of thin shells was developed by Israel \cite{Isr66}
(also see \cite{BerKuzTka} for a detailed derivation and discussion). We choose some hypersurface $\Sigma$, separating the space-time into two regions, `in' and `out', with Gaussian normal coordinates,
\begin{equation}
\label{normal} ds^2 = d\tau^2 - dn^2 - R^2(\tau,n)d\Omega^2 .
\end{equation}
where $\tau$ is the proper time of the observer located at $\Sigma$, and the coordinate $n$ increases from the in-region to the out-region along the outer normal to the hypersurface $\Sigma$.
The hypersurface $\Sigma$, located at  $n=0$, is called a singular shell if some
energy-momentum tensor is localized on it, for example, $T_i^k=S_i^k\delta(n)+\dots$, where $S_i^k$ is the energy-momentum on the shell ($i,k$=0,2,3). In the opposite case, the hypersurface is nonsingular.

From the Einstein equations, we can straightforwardly find sufficiently simple equations of motion and expressions for $S_i^k$. Using them, the accretion of ordinary (non-phantom) shells has been considered in many papers (see, e.g., \cite{BerKuzTka} and the references therein). The accretion of shells, to some extent, can also serve as a model for dark energy accretion.

The case of accretion of a shell with a phantom equation of state was studied in \cite{Beretal05}.
The linear phantom equation of state for the shell has the form $S_0^0=kS_2^2$, where $k>1$. Both finite and infinite motions of the shell are possible. The phantom shell was found to have even more `repulsive' properties in comparison with a shell made of ordinary dark energy with $p<0$. The Schwarzschild geometry surrounded by a phantom shell has the form of a wormhole in all cases except one. In the wormhole geometry, the remote observer cannot see the shell because it is located behind the throat (the Einstein--Rosen bridge).

The back reaction of the accreting matter on the metric can be fully taken into account in the model of thin shells. In the case of a phantom shell, this back reaction can be the critical factor for the formation of global space-time geometry.

In a number of papers, phantom energy was considered as a necessary ingredient for constructing complicated topological structures, like wormholes (variations and generalisations of Einstein--Rosen bridges) \cite{wormh1,wormh2}. In particular, realizing traversable wormholes requires a bridge made of phantom energy between two throats that are asymptotically flat at the
spatial infinity.

\section{Conclusion}

In recent years, the notion of dark energy, in spite of its uncertainty, captured the imagination of cosmology physicists. Dark energy, which in the standard interpretation appears as a substance with negative pressure and provides the observed accelerating expansion of the Universe, represents an extremely important and enigmatic component of the Universe, lying beyond the Standard Model of elementary particles.

Presently, the physical nature of dark energy remains completely unknown, despite many hypotheses offered for its explanation. The modern level of observational technique reveals the presence of dark energy in the Universe only on large scales, in the form of accelerating expansion of the Universe and its effect on the large-scale structure of galaxies and on the CMB anisotropy. Studies of the local interaction of dark energy with black holes, besides being of abstract mathematical interest, seem to be important in the search for new possible appearances of dark energy.

The authors thank the referees, whose remarks allowed significantly improving this review. The work was supported by the RFBR grants 13-02-00257, OFN-17 RAS and NSh-871.2012.2, and the grant FQXi-MGA-1209 from the Foundational Questions Institute.

\end{document}